\documentclass{ws-rmta}

\usepackage{amsmath}
\usepackage{amssymb}
\usepackage{amsxtra}
\usepackage{amsfonts}
\usepackage{bm}
\usepackage{braket}
\usepackage{mathtools}
\usepackage{cite}
\usepackage{longtable}
\usepackage{multirow}
\usepackage{graphicx}
\usepackage{stackrel}
\usepackage{braket}

\usepackage{xcolor}
\definecolor{webgreen}{rgb}{0,.5,0}
\definecolor{webbrown}{rgb}{.6,0,0}
\definecolor{RoyalBlue}{cmyk}{1, 0.50, 0, 0}
\usepackage[colorlinks=true, breaklinks=true, urlcolor=webbrown, linkcolor=RoyalBlue, citecolor=webgreen,backref=page]{hyperref}
\usepackage[cp1251]{inputenc}

\newcommand{\dd}{\,\mathrm{d}}

\DeclareMathOperator{\Tr}{Tr}

\makeatletter

\begin{document}

\markboth{Wei and Witte}{Quantum Interpolating Ensemble}

%\catchline{}{}{}{}{}
%
%%%%%%%%%%%%%%%%%%%%%%%%%%%%%%%%%%%%%%%%%%%%%%%%%%%%%%%%%%%%%%%%%%%%
\title{Quantum Interpolating Ensemble: Biorthogonal Polynomials and Average Entropies}

\author{Lu Wei}

\address{Department of Computer Science, Texas Tech University\\
	Lubbock, Texas 79409, USA\\
	\email{luwei@ttu.edu}}

\author{Nicholas Witte}

\address{School of Mathematics and Statistics, Victoria University of Wellington\\
	Wellington 6012, New Zealand\\
	\email{n.s.witte@protonmail.com} }

\maketitle

%\begin{history}
%\received{(Day Month Year)}
%\revised{(Day Month Year)}
%\end{history}

\begin{abstract}
The density matrix formalism is a fundamental tool in studying various problems in quantum information processing. In the space of density matrices, the most well-known measures are the Hilbert-Schmidt and Bures-Hall ensembles. In this work, the averages of quantum purity and von Neumann entropy for an ensemble that interpolates between these two major ensembles are explicitly calculated for finite-dimensional systems. The proposed interpolating ensemble is a specialization of the $\theta$-deformed Cauchy-Laguerre two-matrix model and new results for this latter ensemble are given in full generality, including the recurrence relations satisfied by their associated bi-orthogonal polynomials when $\theta$ assumes positive integer values.
\end{abstract}

\keywords{quantum entanglement; entanglement entropy; random matrix theory; bi-orthogonal polynomials; interpolating ensemble.}

\ccode{Mathematics Subject Classification 2020: 81P40, 94A17, 15B52, 42C05, 33C45}

\section{Introduction and Quantum Interpolating Ensemble}\label{sec:intro}
Quantum information theory is based on probabilistic interpretations of quantum states to explain various quantum effects. The density matrix formalism introduced by von Neumann~\cite{vN27} provides a natural framework to describe density matrices of quantum states. The density matrix is a fundamental object that encodes all the information of a quantum state. Among the different measures of density matrices, the most well-known and physically relevant ones~\cite{BZ17} are the Hilbert-Schmidt ensemble and the Bures-Hall ensemble.

The Hilbert-Schmidt measure is formulated as follows. Consider a bipartite quantum system consisting of two subsystems $A$ and $B$ in the Hilbert space $\mathcal{H}_{m}$ and $\mathcal{H}_{n}$ (with $m\leq n$), respectively. A random pure state $\Ket{\psi}$, defined as a linear combination of the complete basis of the subsystems, belongs to the composite Hilbert space $\Ket{\psi}\in\mathcal{H}_{m}\otimes\mathcal{H}_{n}$. The reduced density matrix is obtained by partial tracing over the larger system of the full density matrix $\rho=\Ket{\psi}\Bra{\psi}$ as $\rho_{A}=\Tr_{B}\rho$. The resulting density of eigenvalues of $\rho_{A}$ is the Hilbert-Schmidt ensemble~\cite{BZ17}
\begin{equation}\label{eq:HS}
f_{\rm{HS}}\left(\bm{\lambda}\right)\propto
\delta\left(1-\sum_{i=1}^{m}\lambda_{i}\right)\prod_{1\leq i<j\leq m}\left(\lambda_{i}-\lambda_{j}\right)^{2}\prod_{i=1}^{m}\lambda_{i}^{n-m},
\end{equation}
where $\delta(\cdot)$ is the Dirac delta function. The joint density~(\ref{eq:HS}) is also referred to as the Hilbert-Schmidt (random matrix) ensemble, which is the eigenvalue density of a normalized Wishart matrix~\cite{ZPNC11,Osipov10}
\begin{equation}\label{eq:MHS}
\frac{\mathbf{GG}^{\dag}}{\Tr\left(\mathbf{GG}^{\dag}\right)} ,
\end{equation}
with $\mathbf{G}$ being an $m\times n$ complex Gaussian matrix. For the Bures-Hall ensemble, its random pure state $\Ket{\varphi}$ is given by a superposition of the random pure state $\Ket{\psi}$ of the Hilbert-Schmidt measure as $\Ket{\varphi}\propto\Ket{\psi}+\left(\mathbf{U}\otimes I_{n}\right)\Ket{\psi}$, where $\mathbf{U}$ is an $m\times m$ unitary matrix with the measure proportional to $\det\left(I_{m}+\mathbf{U}\right)^{2(n-m)}$. The resulting density of eigenvalues of the reduced density matrix $\rho_{A}=\Tr_{B}\Ket{\varphi}\Bra{\varphi}$ is the (generalized) Bures-Hall ensemble~\cite{Hall98,Zyczkowski01}
\begin{equation}\label{eq:BH}
f_{\rm{BH}}\left(\bm{\lambda}\right) \propto \delta\left(1-\sum_{i=1}^{m}\lambda_{i}\right)
\prod_{1\leq i<j\leq m}\frac{\left(\lambda_{i}-\lambda_{j}\right)^{2}}{\lambda_{i}+\lambda_{j}}\prod_{i=1}^{m}\lambda_{i}^{n-m-\frac{1}{2}}.
\end{equation}
In random matrix theory, the Bures-Hall ensemble~(\ref{eq:BH}) is understood as the joint eigenvalue density of the normalized product of the matrix $I_{m}+\mathbf{U}$ with a complex Gaussian matrix $\mathbf{G}$ as~\cite{ZPNC11,Osipov10}
\begin{equation}	\frac{(I_{m}+\mathbf{U})\mathbf{GG}^{\dag}(I_{m}+\mathbf{U}^{\dag})}{\Tr\left((I_{m}+\mathbf{U})\mathbf{GG}^{\dag}(I_{m}+\mathbf{U}^{\dag})\right)}.
\end{equation}
The Hilbert-Schmidt ensemble~(\ref{eq:HS}) and the Bures-Hall ensemble~(\ref{eq:BH}) are supported in the probability simplex
\begin{equation}\label{eq:sp}
\Lambda=\Bigg\{0\leq\lambda_{m}<\ldots<\lambda_{1}\leq1,~~\sum_{i=1}^{m}\lambda_{i}=1\Bigg\},
\end{equation}
which reflects the constraint $\Tr\rho_{A}=1$ of density matrices. Note also that the normalization constants in the densities~(\ref{eq:HS}) and~(\ref{eq:BH}) are omitted.

The study of the Hilbert-Schmidt measure has received substantial attention, see, for example, the results in~\cite{Lubkin78,Giraud07,Page93,Foong94,Ruiz95,VPO16,Wei17,Wei20,HWC21,Malacarne02,Wei19,ZPNC11,Nadal11,Chen10,Liu18,Hayden06,Aubrun14,Szarek05,Kendon02,Aubrun12}.
These results include information-theoretic studies of different entanglement entropies~\cite{Lubkin78,Giraud07,Page93,Foong94,Ruiz95,VPO16,Wei17,Wei20,HWC21,Malacarne02,Wei19,ZPNC11,Nadal11,Chen10,Liu18,Hayden06,Aubrun14} as well as applications to quantum information processing~\cite{Szarek05,Kendon02,Aubrun12}.
The relatively less-studied Bures-Hall ensemble~\cite{Hall98,Zyczkowski01,Sommers03,Sommers04,Osipov10,Borot12,Zski05,Ye09} has gained renewed interest very recently~\cite{Hu17,Slater19,Sarkar21,Sarkar19,Wei20b,Wei20c,LW21}. This is partially due to the recent breakthrough in probability theory in understanding various aspects of the Bures-Hall ensemble~\cite{BGS09,BGS10,BGS14,FK16,LL19}. Despite the distinct behavior of Hilbert-Schmidt ensemble and Bures-Hall ensemble, an interesting question is whether one could propose an ensemble that interpolates between these two. This question has also been motivated by the observation in~\cite{Sarkar19} that the Bures-Hall ensemble tends to be more conservative than the Hilbert-Schmidt ensemble in estimating entanglement entropies. Namely, the Bures-Hall ensemble leads towards an estimate of less entangled states than the Hilbert-Schmidt ensemble does. In this context, one tries to control the appropriate amount of entanglement as a resource for quantum information processing by constructing new measures that interpolates between the two major ensembles.

In this work, we consider the following ensemble\footnote{On the level of matrix models, a related ensemble has been discussed in~\cite[Eq.~(24)]{Osipov10}.}, also supported in~(\ref{eq:sp}),
\begin{equation}\label{eq:QI}
f\left(\bm{\lambda}\right) = \frac{1}{C}~\delta\left(1-\sum_{i=1}^{m}\lambda_{i}\right)\prod_{1\leq i<j\leq m}\frac{\lambda_{i}-\lambda_{j}}{\lambda_{i}+\lambda_{j}}\left(\lambda_{i}^{\theta}-\lambda_{j}^{\theta}\right)\prod_{i=1}^{m}\lambda_{i}^{a} ,
\end{equation}
termed the quantum interpolating ensemble, where $\theta$ is assumed to be a positive real parameter and $a>-1$. Clearly, the proposed ensemble~(\ref{eq:QI}) reduces to the Hilbert-Schmidt ensemble and the Bures-Hall ensemble as special cases,
\begin{equation}\label{cases}
f\left(\bm{\lambda}\right)
= \begin{cases}
f_{\rm{BH}}\left(\bm{\lambda}\right) & \text{for $\theta=1,~a=n-m-\frac{1}{2}$} \\
f_{\rm{HS}}\left(\bm{\lambda}\right) & \text{for $\theta=2,~a=n-m$}
\end{cases}.
\end{equation}
Namely, as $\theta$ varies from $\theta=1$ to $\theta=2$, the quantum interpolating ensemble interpolates between the Bures-Hall ensemble and the Hilbert-Schmidt ensemble. Due to Schur's Pfaffian identity~\cite{Mehta,Forrester}
\begin{equation}
\prod_{1\leq i<j\leq 2m}\frac{\lambda_{i}-\lambda_{j}}{\lambda_{i}+\lambda_{j}}={\rm{Pf}}\left(\frac{\lambda_{i}-\lambda_{j}}{\lambda_{i}+\lambda_{j}}\right)_{1\leq i,j\leq 2m},
\end{equation}
the proposed ensemble~(\ref{eq:QI}) is described by a Pfaffian point process for any $\theta>0$ except for the special value $\theta=2$ when the ensemble becomes a determinantal point process. Therefore, in the interested interval $\theta\in[1,2]$ the new ensemble~(\ref{eq:QI}) corresponds to the transition from a Pfaffian point process to a determinantal point process. It is also worth mentioning that besides the half integer values of $a$ for the Bures-Hall ensemble and the integer values of $a$ for the Hilbert-Schmidt ensemble, the proposed ensemble is valid for any $a>-1$. Therefore, in addition to $\theta$, the parameter $a$ can be also considered as a deformation parameter that defines the interpolating ensemble.

With an interpolating ensemble being identified, a natural question is what will be the statistical behavior of entanglement entropies over such an ensemble? Will the values of entropies also interpolate between those of the Hilbert-Schmidt ensemble and the Bures-Hall ensemble? Before addressing these information theoretic questions in Section~\ref{IE_applications}, we first study, in a general form, key mathematical aspects of the underlying $\theta$-deformed two-matrix model and the associated bi-orthogonal system, in Section~\ref{CL-two-matrix-ensemble}. The study includes the discovery of new structures of the resulting bi-moment matrix for any $\theta$ that give rise to recurrence relations of the bi-orthogonal polynomials for integer $\theta$, generalizing a few known results in the literature.

\section{The $\theta$-deformed Cauchy-Laguerre Two-matrix Model and Bi-orthogonal Systems}\label{CL-two-matrix-ensemble}
Instead of directly working with the interpolating ensemble of a Pfaffian point process, we will proceed indirectly via the underlying $\theta$-deformed Cauchy-Laguerre two-matrix ensemble. The latter is more conveniently represented by a determinantal point process with the corresponding correlation functions and bi-orthogonal polynomials given explicitly. In particular, we will present new results on the $\theta$-deformed bi-orthogonal system in this section, both for general $\theta>0$ and the specific case of $\theta\in \mathbb{N}$, which fill some gaps in our understanding of the system.

\subsection{General $\theta>0$ case}
We first introduce the two-matrix model, the joint eigenvalue density of which is expressed in terms of two sets of real, positive eigenvalues $\{x_1, \ldots, x_n\}$ and $\{y_1, \ldots, y_n\}$ by the formula~\cite{FL19}
\begin{eqnarray}
&&p(x_1, \ldots, x_n;y_1, \ldots, y_n)\nonumber\\
&=&\frac{1}{(n!)^2Z_n} \prod_{j=1}^{n}x_j^a e^{-x_j} \prod_{k=1}^{n}y_k^b e^{-y_k}
\frac{\prod_{1\leq j<k \leq n}(x_k-x_j)(y_k-y_j)}{\prod_{j,k=1}^{n}(x_j+y_k)}
\prod_{1\leq j<k \leq n}\left(x^{\theta}_k-x^{\theta}_j\right)\left(y^{\theta}_k-y^{\theta}_j\right)\nonumber\\
&=&\frac{1}{(n!)^2Z_n} \prod_{j=1}^{n}x_j^a e^{-x_j} \prod_{k=1}^{n}y_k^b e^{-y_k}
\underset{j,k=1,\ldots,n}{\det} \left( \frac{1}{x_j+y_k}\right)
\underset{\substack{1 \leq j \leq n \\ 0 \leq k \leq n-1}}{\det}\left(x_j^{k\theta}\right)
\underset{\substack{1 \leq j \leq n \\ 0 \leq k \leq n-1}}{\det}\left(y_j^{k\theta}\right),
\end{eqnarray}
for $\Re(\theta)>0$ and all $\Re(a,b)>-1$. The specialization
\begin{equation}\label{eq:ba1}
b=a+1
\end{equation}
of the above density leads to the desired two-matrix model of the unconstrained interpolating ensemble~(\ref{eq:uQI}). It turns out that the general case $\Re(a,b)>-1$ can be treated as conveniently as this special case $b=a+1$, we will therefore consider the former case here in Section~\ref{CL-two-matrix-ensemble}. The normalization of the joint density function is $\int_{\mathbb{R}_{+}^{n}\times\mathbb{R}_{+}^{n}}\dd^{n}x\dd^{n}y\, p(x;y)=1$ through $Z_n(a,b;\theta)$. Henceforth, we will use the abbreviated and commonly used notation
\begin{equation}\label{eq:betaDef}
\beta=\frac{a+b+1}{\theta}=\alpha+1.
\end{equation}

Associated to the $\theta$-deformed Cauchy-Laguerre two-matrix model is the following bi-variate density function
\begin{equation}\label{eq:wb}
w(x,y)=\frac{x^{a}y^{b}e^{-x-y}}{x+y} ,\quad x,y \in [0,\infty).
\end{equation}
One can define an inner product over polynomial spaces $ \cup_{n\geq 0}\Pi_{n}[x] $ using the weight \eqref{eq:wb}.
Let $ f,g \in \cup_{n\geq 0}\Pi_{n}[x] $ then
\begin{equation}\label{IP}
	\langle f,g \rangle := \int_{\mathbb{R}_{+}^{2}} {\dd}x{\dd}y \frac{e^{-x-y}}{x+y}x^{a}y^{b}f(x^{\theta})g(y^{\theta}) .
\end{equation}
The $\theta$-deformed bi-orthogonal systems with respect to the above weight function
are two sequences of {\it normalized bi-orthogonal polynomials} $\{P_n(x),Q_n(y)\}^{\infty}_{n=0}$ satisfying the orthogonality relation
\begin{equation}\label{BOPS_norm}
	\langle P_{m},Q_{n} \rangle = \delta_{m,n} .
\end{equation}
The {\it monic system} $\{\mathcal{P}_n(x),\mathcal{Q}_n(y)\}^{\infty}_{n=0}$ is related via
\begin{align}\label{monic_PQ}
	P_n(x) & = \frac{1}{\sqrt{h_n}} \mathcal{P}_n(x),
\\
	Q_n(y) & = \frac{1}{\sqrt{h_n}} \mathcal{Q}_n(y),
\end{align}
where the normalization constant is
\begin{equation}\label{norm}
	h_n := \langle \mathcal{P}_n,\mathcal{Q}_n \rangle
		= \theta^{-1} \frac{(n!)^2\Gamma(n+\beta)^2}{\Gamma(2n+\beta)\Gamma(2n+\beta+1)} \Gamma(a+1+\theta n)\Gamma(b+1+\theta n) .
\end{equation}
In addition, it will be seen that simpler forms for our results can be obtained if expressed in terms of a third system,
the {\it hybrid} polynomials $\{\tilde{p}_n(x),\tilde{q}_n(y)\}^{\infty}_{n=0}$
\begin{align}\label{}
	\mathcal{P}_n(x) & = (-1)^n n!\frac{\Gamma(a+1+\theta n)\Gamma(n+\beta)}{\Gamma(2n+\beta)}\tilde{p}_n(x) ,
\\
	\mathcal{Q}_n(y) & = (-1)^n n!\frac{\Gamma(b+1+\theta n)\Gamma(n+\beta)}{\Gamma(2n+\beta)}\tilde{q}_n(y) .
\end{align}

The resulting finite bi-moment matrix or Gram matrix is defined by
\begin{equation}\label{I_defn}
	\bm{I} := \left( I_{j,k} \right)^{n-1}_{j,k=0}, \quad
	I_{j,k} := \langle x^{j},y^{k} \rangle, \quad j,k \in \mathbb{Z}_{\geq 0} .
\end{equation}
The bi-moment elements have evaluation, by Eq.~(2.4) Lemma 2.3 of \cite{FL19},
\begin{equation}\label{bi-mom_eval}
	I_{j,k}(a,b;\theta) = \frac{\Gamma(a+1+\theta j)\Gamma(b+1+\theta k)}{a+b+1+\theta(j+k)} = I_{k,j}(b,a;\theta), \quad j,k=0,1,\ldots .
\end{equation}
It is a basic result that the normalization of the joint density function can be expressed as the determinant of the bi-moment matrix,
see Lemma 2.1, Eq.~(2.1) of \cite{FL19}
\begin{equation}
	Z_n(a,b;\theta) = \underset{\substack{0 \leq j \leq n-1 \\ 0 \leq k \leq n-1}}{\det} \left( I_{j,k} \right) .	
\label{det_Z}
\end{equation}
This can also be evaluated, using a variety of methods such as the Cauchy double-alternant formula, and is given by
\begin{equation}
	Z_n(a,b;\theta) =
	\theta^{-n} \left(\prod_{j=1}^{n-1}j!\right)^2 \prod_{j=1}^{n}\frac{\Gamma(j+\beta-1)}{\Gamma(j+n+\beta-1)}
	\prod_{j=0}^{n-1}\Gamma(a+1+\theta j) \prod_{k=0}^{n-1}\Gamma(b+1+\theta k) .
\label{eval_Z}
\end{equation}
In addition to the determinantal formula \eqref{det_Z}, it becomes clear that every aspect of the bi-orthogonal system admits
determinantal representations involving bordered bi-moment matrices. Our first example is also a known result, see Remark 2.7 of \cite{FL19},
\begin{align}
	\mathcal{P}_n(x) & = \frac{1}{Z_n} \det \left(
	\begin{array}{cccc}
	I_{0,0} & \ldots & I_{0,n-1} & 1 \cr
	\vdots  & \ldots & \vdots    & \vdots \cr
	I_{n,0} & \ldots & I_{n,n-1} & x^n
	\end{array} \right),
\label{border_hybrid-p}
\\	
	\mathcal{Q}_n(y) & = \frac{1}{Z_n} \det \left(
	\begin{array}{ccc}
	I_{0,0} & \ldots & I_{0,n} \cr
	\vdots  & \ldots & \vdots    \cr
	I_{n-1,0} & \ldots & I_{n-1,n} \cr
	1         & \ldots & y^n
	\end{array} \right) .
\label{border_hybrid-q}
\end{align}
Using the Cauchy double-alternant formula, one can compute the explicit expansion of the hybrid polynomials in the monomial basis, see Proposition 2.6 of \cite{FL19},
\begin{align}
	\tilde{p}_n(x;a,b;\theta) & = \sum_{l=0}^{n} \frac{(-x)^l}{l!(n-l)!} \frac{\Gamma(n+l+\beta)}{\Gamma(l+\beta)\Gamma(a+1+\theta l)} ,
\label{hybridp}
\\
	\tilde{q}_n(y;a,b;\theta) & = \sum_{l=0}^{n} \frac{(-y)^l}{l!(n-l)!} \frac{\Gamma(n+l+\beta)}{\Gamma(l+\beta)\Gamma(b+1+\theta l)} = \tilde{p}_n(y;b,a;\theta) .
\label{hybridq}
\end{align}

Now denote the column vectors of monomials $\bm{x} = (x^k)_{k\geq 0}$, $\bm{y} = (y^k)_{k\geq 0}$\footnote{Whether these vectors are finite or semi-infinite depends on the context.} and the vectors of normalized bi-orthogonal polynomials $\bm{P} = (P_k(x))_{k\geq 0}$, $\bm{Q} = (Q_k(y))_{k\geq 0}$. These basis vectors are related by lower triangular matrices $\bm{S}_P, \bm{S}_Q$
\begin{equation}\label{LU_poly}
\bm{P} = \bm{S}_P \bm{x}, \qquad \bm{Q} = \bm{S}_Q \bm{y} ,
\end{equation}
which have explicit entries as implied by \eqref{hybridp}, \eqref{hybridq}.
From \eqref{LU_poly} we have the L-U decomposition of the bi-moment matrix
\begin{equation}\label{LU_bimoment}
	\mathbf{I} = \bm{S}_P^{-1} \left( \bm{S}_Q^{-1} \right)^{T} = \left( \bm{S}_Q^{T}\bm{S}_P \right)^{-1} .
\end{equation}

Thinking of the action of a multiplication operator on the normalized polynomial basis, it is clear that it can be written generally as
\begin{equation}\label{Hessenberg}
	x \bm{P} = \bm{X} \bm{P}, \qquad y \bm{Q}^{T} = \bm{Q}^{T} \bm{Y}^{T} ,
\end{equation}
for some lower Hessenberg multiplication matrices $ \bm{X}, \bm{Y} $.
Taking the bi-moment matrix as a semi-infinite matrix $ \mathbf{I}:=(I_{j,k})_{j,k\geq 0} $ and defining the shift matrix $ \mathbf{\Lambda}:=(\delta_{j+1,k})_{j,k\geq 0} $
similarly we can relate some of the notions we have already introduced.
Firstly, we note that the monomial bases are the right and left eigenvectors of the shift matrix and its transpose
\begin{equation}
	\mathbf{\Lambda} \bm{x} = x \bm{x}, \qquad \bm{y}^{T} \mathbf{\Lambda}^{T} = y \bm{y}^{T} .
\label{Shift_eigenvectors}
\end{equation}
Therefore, the multiplication matrices $ \bm{X}, \bm{Y} $ admit the L-U decomposition
\begin{equation}
	\bm{X} = \bm{S}_P \bm{\Lambda} \bm{S}_P^{-1}, \qquad \bm{Y} = \bm{S}_Q \bm{\Lambda} \bm{S}_Q^{-1} .
\label{LU_XY}
\end{equation}

We now provide explicit evaluations of the multiplication matrices $ \bm{X}, \bm{Y} $ for a generic $\theta$.
\begin{proposition}
Let $\Re(a,b)>-1 $ and $\Re(\theta)>0 $. For $ n \geq 0 $, $ 0 \leq m \leq n+1 $, the lower Hessenberg matrix $ \bm{X} $,
in the monic basis, has non-zero elements
\begin{eqnarray}\label{X}
X_{n,m}(a,b)&=&(-1)^{n}\frac{\Gamma(a+1+\theta n)\Gamma(n+\beta)\Gamma(2m+1+\beta)}{\Gamma(a+1+\theta m)\Gamma(m+\beta)\Gamma(2n+\beta)}\nonumber\\
&&\times\sum_{r=m-1}^{n}(-1)^{r}(r+\beta)\binom{n}{r}\binom{r+1}{r+1-m}\frac{\Gamma(n+r+\beta)\Gamma(a+1+\theta(r+1))}{\Gamma(m+2+r+\beta)\Gamma(a+1+\theta r)},
\end{eqnarray}
and the lower Hessenberg matrix $ \bm{Y} $ has elements given by $ Y_{n,m}(a,b) = X_{n,m}(b,a) $.
Note that $X_{n,n+1}=1$ and $X_{n,m}=0$ for any $m>n+1$.
\end{proposition}
\begin{proof}
Writing out the components of \eqref{Hessenberg} and by using orthogonality, one has
\begin{equation}\label{Xint}
	h_{m}X_{n,m} = \int_{\mathbb{R}_{+}^{2}} {\dd}x{\dd}y \frac{e^{-x-y}}{x+y}x^{a+\theta}y^{b}\mathcal{P}_n(x^{\theta})\mathcal{Q}_m(y^{\theta}) .
\end{equation}
We seek to evaluate this using the bordered determinant (see Remark 2.7 of \cite{FL19}) for $ \mathcal{Q}_m $ first and then expand $ \mathcal{P}_n $ afterwards. Thus, we find the right-hand-side of~\eqref{Xint} is given by
\begin{equation}
	\frac{1}{Z_m}\det\left(
	\begin{array}{ccc}
	\ldots & I_{0,l}   & \dots \\
	& \vdots    &       \\
	\ldots & I_{m-1,l} & \dots \\
	\ldots & \int\dd\mu(x,y) x^{\theta}\mathcal{P}_n(x^{\theta})y^{\theta l} & \dots
	\end{array}\right)_{l=0,...,m} ,
\end{equation}
where $\dd\mu(x,y)=w(x,y){\dd}x{\dd}y$. The relevant integral has the evaluation
\begin{equation}
\int\dd\mu(x,y) x^{\theta}\mathcal{P}_n(x^{\theta})y^{\theta l}
=\sum^{n}_{r=0}(-1)^{n-r}\binom{n}{r}\frac{\Gamma(a+1+\theta n)\Gamma(n+\beta)\Gamma(n+r+\beta)}{\Gamma(a+1+\theta r)\Gamma(r+\beta)\Gamma(2n+\beta)} I_{r+1,l} .
\end{equation}
Expanding the determinant along the last row and using the fact that the above $r$-sum is independent of the column index $l$ except for the last factor, we now require the evaluation of the determinant
\begin{equation}
	\frac{1}{Z_m}\det\left(
	\begin{array}{ccc}
	\ldots & I_{0,l}   & \dots \\
	& \vdots    &       \\
	\ldots & I_{m-1,l} & \dots \\
	\ldots & I_{r+1,l} & \dots
	\end{array}\right)_{l=0,...,m} .
\end{equation}
This determinant has the same structure as the standard bi-moment determinant except the last row is out of sequence with the first $m$ rows.
Therefore, it is evaluated as
\begin{equation}
	\prod_{s=0}^{m}\Gamma(b+1+\theta s) \prod_{t=0}^{m-1}\Gamma(a+1+\theta t)\Gamma(a+1+\theta(r+1))
	\det\left(\frac{1}{A_k+B_l}\right)_{k,l=0,...,m} ,
\end{equation}
where $ A_k=a+\frac{1}{2}+\theta k, k=0,...,m-1 $, $ A_{m}=a+\frac{1}{2}+\theta(r+1) $ and $ B_l= b+\frac{1}{2}+\theta l, l=0,...,m $.
We thus arrive at a standard Cauchy double alternant form, which can be computed using
\begin{align}
	\prod_{0\leq k<l \leq m}(A_k-A_l) & = \theta^{m(m+1)/2} (r+2-m)_{m} \prod_{l=1}^{m-1}l! ,
	\\
	\prod_{0\leq k<l \leq m}(B_k-B_l) & = \theta^{m(m+1)/2} \prod_{l=1}^{m}l! ,
	\\
	\prod_{k,l=0}^{m}(A_k+B_l) & = \theta^{(m+1)^2} \frac{(r+m+1+\beta)!}{(r+\beta)!}\prod_{l=1}^{m+1}\frac{(m+l-2+\beta)!}{(l-2+\beta)!} .
\end{align}
Lastly, using the normalization \eqref{eval_Z} we deduce \eqref{X} after some simplification.
\end{proof}

This concludes our discussion of the Cauchy-Laguerre bi-orthogonal system for generic values of $ \theta $. We now move on to the specialization of positive integer valued $\theta$, where the recurrence relations can be explicitly deduced.

\subsection{Integer $\theta$ case}
Our main task here is to elucidate some structures that apply to the Cauchy-Laguerre bi-orthogonal system when one generalizes from $\theta=1$ to arbitrary positive integers. Such structures are not expected to hold in the generic case $\Re(\theta)>0 $ but are useful in many applications including the considered one in quantum information theory. We give a proof of finite order recurrence relations for the general bi-orthogonal system $\{\tilde{p}_n,\tilde{q}_n\}^{\infty}_{n=0}$ for all positive integers $\theta \in \mathbb{N}$ with respect to the general weight function \eqref{eq:wb}. In addition, we will provide explicit examples for $\theta=1, 2$.

The essence of our proof is a generalization of the approach employed in \cite{BGS10} from the rank one shift condition to a rank-$\theta$ condition. This first result is an identity that applies to the bi-moment matrix $ \bm{I} $, and its corresponding consequences for the multiplication matrices $ \bm{X}, \bm{Y} $.
\begin{lemma}
Let $\Re(a,b) > -1$ and $ \theta \in \mathbb{N} $. Furthermore, define the semi-infinite column vectors
	$ \bm{\alpha}_s = (\Gamma(a+\theta-s+\theta k))_{k\geq 0} $, $ \bm{\beta}_s = (\Gamma(b+1+s+\theta k))_{k\geq 0} $
and $ \bm{\pi}_s := \bm{S}_P \bm{\alpha}_s $, $ \bm{\eta}_s := \bm{S}_Q \bm{\beta}_s $, $s\in\mathbb{N}_0$.
The multiplication matrices satisfy the following rank-$\theta$ decomposition
\begin{equation}\label{XY_Id}
	\bm{X}-e^{\pi i\theta} \bm{Y}^{T} = \sum_{s=0}^{\theta-1} (-1)^s \bm{\pi}_s \bm{\eta}_s^{T} .
\end{equation}
\end{lemma}
\begin{proof}
A key observation on the bi-moment evaluation \eqref{bi-mom_eval} is that $x+y$ divides $x^{\theta}-e^{\pi i\theta}y^{\theta}$ without remainder when $ \theta \in \mathbb{N} $
and therefore
\begin{equation}\label{}
	I_{k+1,l}-e^{\pi i\theta}I_{k,l+1} = \sum_{s=0}^{\theta-1} (-1)^s \Gamma(a-s+\theta(k+1))\Gamma(b+1+s+\theta l) .
\end{equation}
Employing semi-infinite matrices $ \mathbf{I}:=(I_{k,l})_{k,l\geq 0} $ and the shift matrix $ \mathbf{\Lambda}:=(\delta_{k+1,l})_{k,l\geq 0}$,
this is written as the rank-$\theta$ decomposition
\begin{equation}
	\mathbf{\Lambda} \mathbf{I}-e^{\pi i\theta} \mathbf{I} \mathbf{\Lambda}^{T} = \sum_{s=0}^{\theta-1} (-1)^s \bm{\alpha}_s \bm{\beta}_s^{T} .
\label{bimoment-Id}
\end{equation}
Upon premultiplying \eqref{bimoment-Id} by $\bm{S}_P$ and postmultiplying by $\bm{S}_Q^{T}$,
and recalling the L-U decomposition of the bi-moment matrix \eqref{LU_bimoment} as well as the multiplication matrices $ \bm{X}, \bm{Y} $ \eqref{LU_XY},
we deduce \eqref{XY_Id}.
\end{proof}

In order to proceed further towards the recurrence relations, it is necessary to construct rank-$\theta$ annihilators, along the lines that was done in \cite{BGS10} for $\theta=1$, for the right-hand side of \eqref{XY_Id}. This can be achieved recursively in $ \theta $ steps, however we show the explicit details for $ \theta=1,2 $ only.
\begin{theorem}\label{recurrence-one}
For $\Re(a,b) > -1$ and $\theta=1$, the hybrid polynomials $\tilde{p}_n(x),\tilde{q}_n(x)$ satisfy the third order recurrence relations
\begin{equation}\label{RRp-one}
x\left(a_{n,1}\tilde{p}_{n+1}(x)+a_{n,0}\tilde{p}_{n}(x)\right)\\
=r_{n,2}\tilde{p}_{n+2}(x)+r_{n,1}\tilde{p}_{n+1}(x)+r_{n,0}\tilde{p}_{n}(x)+r_{n,-1}\tilde{p}_{n-1}(x) ,
\end{equation}
and
\begin{equation}\label{RRq-one}
y\left(a_{n,1}\tilde{q}_{n+1}(y)+a_{n,0}\tilde{q}_{n}(y)\right)\\
=s_{n,2}\tilde{q}_{n+2}(y)+s_{n,1}\tilde{q}_{n+1}(y)+s_{n,0}\tilde{q}_{n}(y)+s_{n,-1}\tilde{q}_{n-1}(y) ,
\end{equation}
where
\begin{equation}\label{RRa-one}
	a_{n,1} = a_{n,0} = 1
\end{equation}
and the coefficients $r_{n}(a,b) $ are given by
\begin{eqnarray}
r_{n,2}&=&-\frac{(n+2)(a+n+2)(\beta+n+1)}{(\beta+2n+2)(\beta+2n+3)},\label{RRr-one:2} \\
r_{n,1}&=&\frac{(n+2)(a+n+2)(\beta +n+1)}{\beta+2n+3}-\frac{(n+1)(a+n+1)(\beta +n)}{\beta+2n},\label{RRr-one:1} \\
r_{n,0}&=&(n+2)(a+n+2)(\beta+n+1)\left(\frac{1}{\beta+2 n+2}-\frac{1}{2}\right)+(n+1)(a+n+1)(\beta+n)\nonumber \\
&&-n(a+n)(\beta+n-1)\left(\frac{1}{\beta+2n-1}+\frac{1}{2}\right),\label{RRr-one:0} \\
r_{n,-1}&=&\frac{n(b+n)(\beta+n-1)}{(\beta+2n-1)(\beta+2n)}, \label{RRr-one:-1} \\
\end{eqnarray}
and $s_{n,k}(a,b)=r_{n,k}(b,a)$.
\end{theorem}

\begin{theorem}\label{recurrence-two}
For $\Re(a,b) > -1$ and $\theta=2$, the hybrid polynomials $\tilde{p}_n(x),\tilde{q}_n(x)$ satisfy the fourth order recurrence relations
\begin{eqnarray}\label{RRp-tw0}
&&x\left(a_{n,2}\tilde{p}_{n+2}(x)+a_{n,1}\tilde{p}_{n+1}(x)+a_{n,0}\tilde{p}_{n}(x)\right)\nonumber\\
&=&r_{n,3}\tilde{p}_{n+3}(x)+r_{n,2}\tilde{p}_{n+2}(x)+r_{n,1}\tilde{p}_{n+1}(x)+r_{n,0}\tilde{p}_{n}(x)+r_{n,-1}\tilde{p}_{n-1}(x)
\end{eqnarray}
and
\begin{eqnarray}\label{RRq-tw0}
&&y\left(a_{n,2}\tilde{q}_{n+2}(y)+a_{n,1}\tilde{q}_{n+1}(y)+a_{n,0}\tilde{q}_{n}(y)\right)\nonumber\\
&=&s_{n,3}\tilde{q}_{n+3}(y)+s_{n,2}\tilde{q}_{n+2}(y)+s_{n,1}\tilde{q}_{n+1}(y)+s_{n,0}\tilde{q}_{n}(y)+s_{n,-1}\tilde{q}_{n-1}(y),
\end{eqnarray}
where
\begin{equation}
	a_{n,2} = \frac{1}{2n+3+\beta} ,
	\quad
	a_{n,1} = \frac{2(2n+2+\beta)}{(2n+1+\beta)(2n+3+\beta)} ,
	\quad
	a_{n,0} = \frac{1}{2n+1+\beta} ,
\end{equation}
and the coefficients $r_{n,k}(a,b) $ are given by \eqref{rCoeff:p3} to \eqref{rCoeff:m1} below, and $ s_{n,k}(a,b)=r_{n,k}(b,a) $.
\end{theorem}
\begin{proof}
For any vector $\bm{\pi}$, we construct the semi-infinite diagonal matrix $ \bm{D}_{\bm{\pi}} $ so that $ \bm{\pi} = \bm{D}_{\bm{\pi}}\bm{1} $.
From the knowledge that the unit vector $ \bm{1} $ (or any constant vector) is left-annihilated by $ \bm{\Lambda}-Id $
we can left-annihilate the $ s=0 $ term on the right-hand side of \eqref{XY_Id} by premultiplying with $ \left( \bm{\Lambda}-Id \right)\bm{D}_{\bm{\pi}_0}^{-1}  $.
To annihilate the remaining $s=1$ term we need to calculate $ \left( \bm{\Lambda}-Id \right)\bm{D}_{\bm{\pi}_0}^{-1}\bm{D}_{\bm{\pi}_1}\bm{1} =: \bm{\psi} $.
A simple calculation gives the components of $ \bm{\psi} $ as
\begin{equation}\label{}
	\psi_n = \frac{\pi_{1,n}}{\pi_{0,n}} - \frac{\pi_{1,n+1}}{\pi_{0,n+1}},~n = 0,1, \dots ,
\end{equation}
assuming $ \pi_{0,n} \neq 0 $.
Our required second left-annihilator is therefore $ \left( \bm{\Lambda}-Id \right)\bm{D}_{\bm{\psi}}^{-1} $ and so the composite operator
is the second order difference operator
\begin{equation}\label{}
	\left( \bm{\Lambda}-Id \right)\bm{D}_{\bm{\psi}}^{-1} \left( \bm{\Lambda}-Id \right)\bm{D}_{\bm{\pi}_0}^{-1} .
\end{equation}
This recursion can be repeated up to $ \theta $ levels leaving us with a $\theta$-order difference operator in the general case, modulo the non-vanishing condition given above. Before we compute this our final step in deriving the recurrence relation for $ P_n $ is to put some of these pieces together. Let us act on the first equation of \eqref{Hessenberg} with this operator - doing so on the left-hand side gives
\begin{equation}\label{}
x \left( \bm{\Lambda}-Id \right)\bm{D}_{\bm{\psi}}^{-1} \left( \bm{\Lambda}-Id \right)\bm{D}_{\bm{\pi}_0}^{-1} \bm{P} ,
\end{equation}
whereas acting on the right-hand side gives
\begin{equation}\label{}
	\left( \bm{\Lambda}-Id \right)\bm{D}_{\bm{\psi}}^{-1} \left( \bm{\Lambda}-Id \right)\bm{D}_{\bm{\pi}_0}^{-1} \bm{X} \bm{P} =: \bm{A} \bm{P} ,
\end{equation}
which defines a banded matrix $ \bm{A} $.
This banded matrix has non-zero elements only for $\theta+1$ super-diagonals above the diagonal
- the $\theta$-order difference operator adds $\theta$ super-diagonals to the initial single one of the lower Hessenberg $ \bm{X} $ -
and a single sub-diagonal - this operator does not add any additional sub-diagonals to the upper Hessenberg matrix $ \bm{Y}^{T} $.
The analogous result for $ \bm{Q} $ is
\begin{equation}\label{}
	y\bm{Q}^{T} \bm{D}_{\bm{\eta}_0}^{-1}\left( \bm{\Lambda}^{T}-Id \right) \bm{D}_{\bm{\psi}}^{-1}\left( \bm{\Lambda}^{T}-Id \right) =: \bm{Q}^{T} \bm{B} ,
\end{equation}
where
\begin{equation}\label{}
	\psi_n = - \frac{\eta_{1,n}}{\eta_{0,n}} + \frac{\eta_{1,n+1}}{\eta_{0,n+1}},~n = 0,1, \dots,~\eta_{0,n} \neq 0.
\end{equation}
For $\theta=2$, the second order difference operator acting on $ \bm{P} $ has $n$-th component
\begin{eqnarray}
&&\frac{\pi_{0,n+1}}{\pi_{0,n+2}\pi_{1,n+1}-\pi_{0,n+1}\pi_{1,n+2}} P_{n+2}
+\frac{\pi_{0,n+1}}{\pi_{0,n+1}\pi_{1,n}-\pi_{0,n}\pi_{1,n+1}} P_{n}\nonumber\\
&&+\frac{\pi_{0,n+1}\left(\pi_{0,n}\pi_{1,n+2}-\pi_{0,n+2}\pi_{1,n}\right)}
{\left(\pi_{0,n+1}\pi_{1,n}-\pi_{0,n}\pi_{1,n+1}\right)\left(\pi_{0,n+2}\pi_{1,n+1}-\pi_{0,n+1}\pi_{1,n+2}\right)} P_{n+1}
\label{2ndOrderDO}
\end{eqnarray}
and on $\bm{Q}$,
\begin{eqnarray}
&&-\frac{\eta_{0,n+1}}{\eta_{0,n+2}\eta_{1,n+1}-\eta_{0,n+1}\eta_{1,n+2}} Q_{n+2}
-\frac{\eta_{0,n+1}}{\eta_{0,n+1}\eta_{1,n}-\eta_{0,n}\eta_{1,n+1}} Q_{n}\nonumber\\
&&-\frac{\eta_{0,n+1}\left(\eta_{0,n}\eta_{1,n+2}-\eta_{0,n+2}\eta_{1,n}\right)}
{\left(\eta_{0,n+1}\eta_{1,n}-\eta_{0,n}\eta_{1,n+1}\right)\left(\eta_{0,n+2}\eta_{1,n+1}-\eta_{0,n+1}\eta_{1,n+2}\right)} Q_{n+1}.
\end{eqnarray}
	
In the case at hand we compute the components of $ \bm{\pi}_{0},\bm{\pi}_{1} $ and $ \bm{\eta}_{0},\bm{\eta}_{1} $ to be
\begin{align}
	\pi_{0,n} & = \frac{n!}{\sqrt{h_n}}\frac{\Gamma(a+1+2n)\Gamma(\beta+n)}{\Gamma(\beta+2n)}\left(a+1+2n(n+\beta)\right),
	\\
	\pi_{1,n} & = \frac{n!}{\sqrt{h_n}}\frac{\Gamma(a+1+2n)\Gamma(\beta+n)}{\Gamma(\beta+2n)},
	\\
	\eta_{0,n} & = \frac{n!}{\sqrt{h_n}}\frac{\Gamma(b+1+2n)\Gamma(\beta+n)}{\Gamma(\beta+2n)},
	\\
	\eta_{1,n} & = \frac{n!}{\sqrt{h_n}}\frac{\Gamma(b+1+2n)\Gamma(\beta+n)}{\Gamma(\beta+2n)}\left(b+1+2n(n+\beta)\right).
\end{align}
We thus arrive at the recurrences for our hybrid system \eqref{RRp-tw0} and \eqref{RRq-tw0}. There are two observations to make about these results.
Firstly, the method generates an overall common factor for the $a$ coefficients, which is subsequently present in all the $r$-coefficients,
and our results have this factor removed. For $\theta=2$ it is a $ (-1)^n(a+1+2(n+1)(n+1+\beta)) $ which is non-zero for $\Re(a)>-1$, $\Re(\beta)>-1 $, and $ n\in\mathbb{Z}_{\geq 0} $. Secondly, note that the difference operators defined by the left-hand sides of the recurrences \eqref{RRp-tw0} and \eqref{RRq-tw0} are identical and symmetrical with respect to $a,b$.
	
Finally, the $r_n$, $s_n$ coefficients can be deduced in a number of ways, such as employing the explicit series form for the polynomials and peeling off the leading terms from the highest degree ($n+3$) down in four successive iterations,
or by acting upon the $ \bm{X} $ matrix with the second-order difference operator \eqref{2ndOrderDO}.
Either way we find
\begin{eqnarray}
r_{n,3}&=&-\frac{(n+3)(a+2n+5)(a+2n+6)(\beta+n+2)}{(\beta+2n+3)(\beta+2n+4)(\beta+2n+5)},\label{rCoeff:p3}\\
r_{n,2}&=&\frac{(n+3)(a+2n+5)(a+2n+6)(\beta+n+2)}{(\beta+2n+3)(\beta+2n+5)}\nonumber\\
&&-\frac{(n+2)(a+2n+3)(a+2n+4)(\beta+n+1)}{(\beta+2n+1)(\beta+2n+3)},\label{rCoeff:p2}\\
r_{n,1}&=&-\frac{(n+1)(a+2n+1)(a+2n+2)(\beta+n)(\beta+2n+2)}{2(\beta+2n)(\beta+2n+1)}\nonumber\\
&&+\frac{(n+2) (a+2 n+3) (a+2 n+4) (\beta +n+1) (\beta +2 n+2)}{(\beta +2 n+1) (\beta +2 n+3)}\nonumber\\
&&-\frac{(n+3) (a+2 n+5) (a+2 n+6) (\beta +n+2) (\beta +2 n+2)}{2 (\beta +2 n+3) (\beta +2 n+4)},\label{rCoeff:p1}\\
r_{n,0}&=&\frac{(n+3) (a+2 n+5) (a+2 n+6) (\beta +n+2) (\beta +2 n)}{6 (\beta +2 n+3)}\nonumber\\	
&&-\frac{(n+2) (a+2 n+3) (a+2 n+4) (\beta +n+1) (\beta +2 n)}{\beta +2 n+3} \left(\frac{1}{\beta +2 n+1}+\frac{1}{2}\right)\nonumber\\
&&+\frac{(n+1) (a+2 n+1) (a+2 n+2)(\beta +n)}{\beta +2 n+3} \left(\frac{2 \beta +4 n+3}{\beta +2 n+1}+\frac{1}{2} (\beta +2 n)\right)\nonumber\\
&&-\frac{n (a+2 n-1) (a+2 n) (\beta +n-1)}{\beta +2 n+3}\left(\frac{\beta +2 n+1}{\beta +2 n-1}+\frac{1}{6} (\beta +2 n)\right),\label{rCoeff:p0}\\
r_{n,-1}&=&-\frac{n (b+2 n-1) (b+2 n) (\beta +n-1)}{(\beta +2 n-1) (\beta +2 n) (\beta +2 n+1)}.\label{rCoeff:m1}
\end{eqnarray}
Independently of the $r$-coefficients, the $s$-coefficients were computed and found to verify the symmetry relation $ s_{n,l}(a,b)=r_{n,l}(b,a) $.
\end{proof}

From the workings in the above proof one sees the structures of the recurrence relations in the general case of $\theta\in\mathbb{N}$ as summarized in the following corollary.
\begin{corollary}
Let $ \theta \in \mathbb{N} $ and $\Re(a,b) > -1$.
The hybrid polynomials satisfy the recurrence relations
\begin{equation}
	x \sum_{k=0}^{\theta} a_{n,k}\tilde{p}_{n+k}(x) = \sum_{l=-1}^{\theta+1} r_{n,l}\tilde{p}_{n+l}(x)
\end{equation}
and
\begin{equation}
	y \sum_{k=0}^{\theta} a_{n,k}\tilde{q}_{n+k}(y)	= \sum_{l=-1}^{\theta+1} s_{n,l}\tilde{q}_{n+l}(y) ,
\end{equation}
with $s_{n,l}(a,b)=r_{n,l}(b,a)$.
\end{corollary}
\noindent For any given $\theta\in\mathbb{N}$, explicit expressions of the recurrence coefficients can be obtained in a similar manner as in Theorems~\ref{recurrence-one} and~\ref{recurrence-two} but with increasingly more effort for a larger $\theta$.

In concluding this section we give a special result of interest on the recurrence relation of the $\theta$-deformed bi-orthogonal polynomials~\eqref{eq:p} corresponding to the Hilbert-Schmidt ensemble, i.e., when $\theta=2$ in addition to the specialization~(\ref{eq:ba1}). This result is useful in the computation of higher order moments of entanglement entropies. Namely, upon the specialization of Theorem~\ref{recurrence-two} with $\theta=2$, $b=a+1$ and using the application-wise more convenient notation
\begin{equation}\label{eq:pp}
p_{j}(x)=(-1)^{j}\sqrt{2}\tilde{p}_j(x),
\end{equation}
we arrive at the following corollary.
\begin{corollary}\label{p:3}
The $\theta$-deformed bi-orthogonal polynomials $p_{j}$ in~(\ref{eq:p}) for $\theta=2,~b=a+1$ satisfy the fourth order recurrence relations
\begin{eqnarray}
&&x\left(a_{2}p_{j+2}\left(x\right)+a_{1}p_{j+1}\left(x\right)+a_{0}p_{j}\left(x\right)\right)\nonumber\\
&=&r_{3}p_{j+3}\left(x\right)+r_{2}p_{j+2}\left(x\right)+r_{1}p_{j+1}\left(x\right)+r_{0}p_{j}\left(x\right)+r_{-1}p_{j-1}\left(x\right),
\end{eqnarray}
where the coefficients are explicitly given by
\begin{align}
	a_{2}	&=\frac{2aj+3a+2j^2+6j+5}{a+2j+4} \label{eq:a2}\\
	a_{1}	&=-\frac{2(a+2j+3)\left(2aj+3a+2j^2+6j+5\right)}{(a+2j+2)(a+2j+4)} \label{eq:a1}\\
	a_{0}	&=\frac{2aj+3a+2j^2+6j+5}{a+2j+2} \label{eq:a0}\\
	r_{3}	&=\frac{(j+3)(a+j+3)\left(2aj+3a+2j^2+6j+5\right)}{a+2j+4} \label{eq:r3}\\
	r_{2}	&=\frac{a^3+6a^{2}j+12a^2+12aj^2+46aj+41a+8j^3+46j^2+82j+42}{\left(2aj+3a+2j^2+6j+5\right)^{-1}(a+2j+2)(a+2j+4)} \label{eq:r2}\\
	r_{1}	&=\frac{(a+2j+3)\left(2aj+3a+2j^2+6j+5\right)\left(2a^2+6aj+9a+6j^2+18j+10\right)}{(a+2j+2)(a+2j+4)} \label{eq:r1}\\
	r_{0}	&=\frac{a^3+6a^{2}j+6a^2+12aj^2+26aj+11a+8j^3+26j^2+22j+6}{\left(2aj+3a+2j^2+6j+5\right)^{-1}(a+2j+2)(a+2j+4)} \label{eq:r0}\\
	r_{-1}	&=\frac{j(a+j)\left(2aj+3a+2j^2+6j+5\right)}{a+2j+2}.\label{eq:rm1}
\end{align}
\end{corollary}
The corresponding recurrence relation of the dual polynomial~\eqref{eq:q} can be also similarly found.

\section{Applications to Quantum Information Theory}\label{IE_applications}
In this section, we study a special case, relevant to quantum information theory, of the above discussed bi-orthogonal system that gives rise to the interpolating ensemble of interest. We first outline the corresponding correlation kernels before showing a new result on the kernel factorizations. We then perform analytical and numerical study on the average behavior of entanglement entropies over the interpolating ensemble.

\subsection{Entanglement entropies and correlation kernels}	
For the quantum bipartite system introduced in Section~\ref{sec:intro}, the degree of entanglement of subsystems $A$ and $B$ is estimated by entanglement entropies, which are functions of the eigenvalues (entanglement spectrum) of a given ensemble. Any function that satisfies a list of axioms can be considered as an entanglement entropy. In particular, an entropy should monotonically change from the separable state
\begin{equation}\label{eq:s}
	\lambda_{1}=1,~~\lambda_{2}=\dots=\lambda_{m}=0 ,
\end{equation}
to the maximally-entangled state
\begin{equation}\label{eq:e}
	\lambda_{1}=\lambda_{2}=\dots\lambda_{m}=\frac{1}{m}.
\end{equation}
A standard one we consider here is quantum purity~\cite{BZ17}
\begin{equation}\label{eq:P}
	S_{\rm{P}}=\sum_{i=1}^{m}\lambda_{i}^{2},
\end{equation}
supported in $S_{\rm{P}}\in[1/m,1]$, which attains the separable state and maximally-entangled state when $S_{\rm{P}}=1$ and when $S_{\rm{P}}=1/m$, respectively. Quantum purity~(\ref{eq:P}) is an example of polynomial entropies, whereas a well-known non-polynomial entropy is von Neumann entropy~\cite{BZ17}
\begin{equation}\label{eq:vN}
	S_{\rm{vN}}=-\sum_{i=1}^{m}\lambda_{i}\ln\lambda_{i}.
\end{equation}
The von Neumann entropy~(\ref{eq:vN}) is supported in $S_{\rm{vN}}\in[0,\ln{m}]$ that achieves the separable state and maximally-entangled state when $S_{\rm{vN}}=0$ and when $S_{\rm{vN}}=\ln{m}$, respectively. Statistical information of entanglement entropies is encoded through their moments: the first moment (average value) implies the typical behavior of entanglement, the second moment (variance) specifies the fluctuation around the typical value, and the higher order moments (such as skewness and kurtosis) describe the tails of the distributions. We focus on the average entanglement entropies in this work, whereas the study of higher order moments would make full use of the results derived in Section~\ref{CL-two-matrix-ensemble}.

Moment computation over an ensemble with the probability constraint $\delta\left(1-\sum_{i=1}^{m}\lambda_{i}\right)$ is typically performed over an ensemble without the constraint~\cite{Page93,Ruiz95,Wei17,Wei20,HWC21,Sarkar19,Wei20b,Wei20c,LW21}. As will be seen, the unconstrained version of the interpolating ensemble~(\ref{eq:QI}) is given by
\begin{equation}\label{eq:uQI}
h\left(\bm{x}\right)=\frac{1}{C'}\prod_{1\leq i<j\leq m}\frac{x_{i}-x_{j}}{x_{i}+x_{j}}\left(x_{i}^{\theta}-x_{j}^{\theta}\right)\prod_{i=1}^{m}x_{i}^{a}e^{-x_{i}},
\end{equation}
where $x_{i}\in[0,\infty)$, $i=1,\dots,m$. This ensemble has been recently proposed in~\cite{FL19} in connection to a $\theta$-deformed Cauchy-Laguerre two-matrix model. In the case when $\theta=1$, the corresponding ensembles have been studied in~\cite{BGS09,BGS10,BGS14,FK16,LL19}.
We now move on to the moment relations of entanglement entropies between the proposed ensemble~(\ref{eq:QI}) and its unconstrained version~(\ref{eq:uQI}). Firstly, the density $g_{d}(r)$ of the trace
\begin{equation}\label{eq:tr}
	r=\sum_{i=1}^{m}x_{i},~~~~r\in[0,\infty) ,
\end{equation}
of the unconstrained ensemble~(\ref{eq:uQI}) is obtained as
\begin{align}
	g_{d}(r)&=\int_{\bm{x}}h(\bm{x})\delta\left(r-\sum_{i=1}^{m}x_{i}\right)\prod_{i=1}^{m}\dd x_{i}
\nonumber \\
			&=\frac{C}{C'}e^{-r}r^{d-1}\int_{\bm{\lambda}}f(\bm{\lambda})\prod_{i=1}^{m}\dd\lambda_{i}
\nonumber \\
			&=\frac{1}{\Gamma\left(d\right)}e^{-r}r^{d-1},
\label{eq:g}
\end{align}
where we have used the change of variables
\begin{equation}\label{eq:cv}
	x_{i}=r\lambda_{i},~~~~i=1,\ldots,m ,
\end{equation}
and the resulting Jacobian calculation leads to the normalization $\Gamma(d)$ with
\begin{equation}\label{eq:d}
	d=\frac{m}{2}(m\theta-\theta+2a+2).
\end{equation}
The above calculation implies that the density $h(\bm{x})$ can be factored as
\begin{equation}\label{eq:h2f}
	h(\bm{x})\prod_{i=1}^{m}\dd x_{i}=f(\bm{\lambda})g_{d}(r)\dd r\prod_{i=1}^{m}\dd\lambda_{i},
\end{equation}
i.e., the random variable $r$ is independent of each $\lambda_{i}$ (hence independent of $S_{\rm{P}}$ and $S_{\rm{vN}}$). Similar factorizations also exist for the Hilbert-Schmidt ensemble~\cite{Page93} and the Bures-Hall ensemble~\cite{Sarkar19,Wei20c,LW21}. Introducing the corresponding quantum purity of the unconstrained ensemble
\begin{equation}\label{eq:TP}
	T_{\rm{P}}=\sum_{i=1}^{m}x_{i}^{2},
\end{equation}
the $k$-th moment of quantum purity $S_{\rm{P}}$ is represented as
\begin{align}
	\mathbb{E}_{f}\!\left[S_{\rm{P}}^{k}\right]
		&=\int_{\bm{\lambda}}S_{\rm{P}}^{k}~f(\bm{\lambda})\prod_{i=1}^{m}\dd\lambda_{i}
\nonumber \\
		&=\int_{\bm{\lambda}}\frac{T_{\rm{P}}^{k}}{r^{2k}}f(\bm{\lambda})\prod_{i=1}^{m}\dd\lambda_{i}\int_{r}g_{d+2k}(r)\dd r
\nonumber \\
		&=\frac{\Gamma(d)}{\Gamma(d+2k)}\int_{\bm{\lambda}}\int_{r}T_{\rm{P}}^{k}~f(\bm{\lambda})g_{d}(r)\dd r\prod_{i=1}^{m}\dd\lambda_{i}
\nonumber \\
		&=\frac{\Gamma(d)}{\Gamma(d+2k)}\mathbb{E}_{h}\!\left[T_{\rm{P}}^{k}\right],
\end{align}
where we have used the change of variables~(\ref{eq:cv}) and the independence property~(\ref{eq:h2f}). Therefore, computing the $k$-th moment of $S_{\rm{P}}$ can be converted to computing the $k$-th moment of $T_{\rm{P}}$. In particular, the first moments are related by
\begin{equation}\label{eq:f2hP1}
	\mathbb{E}_{f}\!\left[S_{\rm{P}}\right]=\frac{1}{d(d+1)}\mathbb{E}_{h}\!\left[T_{\rm{P}}\right].
\end{equation}
We now introduce von Neumann entropy of the unconstrained ensemble
\begin{equation}\label{eq:TvN}
	T_{\rm{vN}}=\sum_{i=1}^{m}x_{i}\ln x_{i},
\end{equation}
that leads to the identity
\begin{equation}
	S_{\rm{vN}}=\ln r-r^{-1}T_{\rm{vN}},
\end{equation}
then the first moment relation is similarly obtained as
\begin{align}
	\mathbb{E}_{f}\!\left[S_{\rm{vN}}\right]
		&=\int_{\bm{\lambda}}S_{\rm{vN}}~f(\bm{\lambda})\prod_{i=1}^{m}\dd\lambda_{i}\int_{r}g_{d+1}(r)\dd r
\nonumber \\
		&=\frac{\Gamma(d)}{\Gamma(d+1)}\left(\int_{r}g_{d}(r)r\ln r\dd r-\int_{\bm{\lambda}}\int_{r}T_{\rm{vN}}~f(\bm{\lambda})g_{d}(r)\dd r\prod_{i=1}^{m}\dd\lambda_{i}\right)
\nonumber \\
		&=\psi_{0}(d+1)-\frac{1}{d}\mathbb{E}_{h}\!\left[T_{\rm{vN}}\right],
\label{eq:f2hvN1}
\end{align}
where we have also used
\begin{equation}\label{eq:1eln}
\int_{0}^{\infty}e^{-r}r^{a-1}\ln{r}\dd r=\Gamma(a)\psi_{0}(a),~~~~\Re(a)>0
\end{equation}
with $\psi_{0}(x)=\dd\ln\Gamma(x)/\dd x$ denoting the digamma function~\cite{Prudnikov86}. For a positive integer $l$, the digamma function admits the following useful identities
\begin{subequations}
	\begin{align}
	&\psi_{0}(l)=-\gamma+\sum_{k=1}^{l-1}\frac{1}{k} \\
	&\psi_{0}\left(l+\frac{1}{2}\right)=-\gamma-2\ln2+2\sum_{k=0}^{l-1}\frac{1}{2k+1},
	\end{align}
\end{subequations}
where $\gamma\approx0.5772$ is Euler's constant.

Computing the $k$-th moment of the above defined entropies requires the $k$-point correlation function of the unconstrained ensemble~\eqref{eq:uQI}, which is recently shown to follow a Pfaffian point process of a $2k\times2k$ antisymmetric matrix~\cite{FL19},
\begin{equation}\label{eq:rk}
	\rho_{k}(x_{1},\dots,x_{k})\propto{\rm{Pf}}\left(
	\begin{array}{cc}
	\Delta K_{11}(x_{i},x_{j}) & \Sigma K_{01}(x_{i},x_{j}) \\
	-\Sigma K_{01}(x_{j},x_{i}) & \Delta K_{00}(x_{i},x_{j}) \\
	\end{array}\right)_{1\leq i,j\leq k},
\end{equation}
where the kernels
\begin{subequations}
	\begin{align}
	\Delta K_{00}(x,y)&=K_{00}(x,y)-K_{00}(y,x),
\\
	\Sigma K_{01}(x,y)&=K_{01}(x,y)+K_{10}(y,x),
\\
	\Delta K_{11}(x,y)&=K_{11}(x,y)-K_{11}(y,x),
	\end{align}
\end{subequations}
are written in terms of those of the $\theta$-deformed Cauchy-Laguerre bi-orthogonal ensemble $K_{00}(x,y)$, $K_{01}(x,y)$, $K_{10}(x,y)$, and $K_{11}(x,y)$. As a result, the computation of various statistical quantities of the quantum interpolating ensemble can be performed over those kernels. The kernels can be expressed via the following Fox H-functions~\cite{FL19}
\begin{subequations}
	\begin{align}
	H_{q}(x)&=H_{2,3}^{1,1}\left(\begin{array}{c} (-\alpha-m,1); (m,1) \\ (0,1); (-q,\theta), (-\alpha,1) \end{array}\Big|tx^{\theta}\Big.\right),
\label{eq:H1}\\
	G_{q}(x)&=H_{2,3}^{2,1}\left(\begin{array}{c} (-\alpha-m,1); (m,1) \\ (0,1), (-q,\theta); (-\alpha,1) \end{array}\Big|tx^{\theta}\Big.\right),
\label{eq:H2}
	\end{align}
\end{subequations}
as
\begin{subequations}
	\begin{align}
	K_{00}(x,y)&=\theta\int_{0}^{1}t^{\alpha}H_{a}(x)H_{a+1}(y)\dd t,
\label{eq:K00I}\\
	K_{01}(x,y)&=\theta x^{2a+1}\int_{0}^{1}t^{\alpha}H_{a}(y)G_{a+1}(x)\dd t,
\label{eq:K01I}\\
	K_{10}(x,y)&=\theta y^{2a+1}\int_{0}^{1}t^{\alpha}H_{a+1}(x)G_{a}(y)\dd t,
\label{eq:K10I}\\
	K_{11}(x,y)&=\theta(xy)^{2a+1}\int_{0}^{1}t^{\alpha}G_{a+1}(x)G_{a}(y)\dd t-\frac{x^{a}y^{a+1}}{x+y},
\label{eq:K11I}
	\end{align}
\end{subequations}
where the definition
\begin{equation}\label{eq:alpha}
\alpha=\frac{2(a+1)}{\theta}-1.
\end{equation}
follows from the notation~(\ref{eq:betaDef}). In general, the Fox H-function is defined through the following contour integral~\cite{Mathai10}
\begin{eqnarray}\label{eq:FH}
&&H_{p,q}^{m,n}\left(\begin{array}{c} (a_{1},A_{1}),\ldots,(a_{n},A_{n}); (a_{n+1},A_{n+1}),\ldots, (a_{p},A_{p}) \\ (b_{1},B_{1}),\ldots, (b_{m},B_{m}); (b_{m+1},B_{m+1}),\ldots, (b_{q},B_{q}) \end{array}\Big|x\Big.\right)\\
&=&\frac{1}{2\pi\imath}\int_{\mathcal{L}}{\frac{\prod_{j=1}^m\Gamma\left(b_{j}+B_{j}s\right)\prod_{j=1}^n\Gamma\left(1-a_{j}-A_{j}s\right)}{\prod_{j=n+1}^p \Gamma\left(a_{j}+A_{j}s\right)\prod_{j=m+1}^q\Gamma\left(1-b_{j}-B_{j}s\right)}}x^{-s}\dd s,
\end{eqnarray}
where the contour $\mathcal{L}$ separates the poles of $\Gamma\left(b_{j}+B_{j}s\right)$ from the poles of $\Gamma\left(1-a_{j}-A_{j}s\right)$. In the special case $A_{1}=\dots=A_{p}=1$ and $B_{1}=\dots=B_{q}=1$, the Fox H-function reduces to the Meijer G-function~\cite{Mathai10}. The integral forms of the kernel functions~(\ref{eq:K00I})--(\ref{eq:K11I}) are useful for the computation of mean entropies, whereas several other distinct representations of the kernel functions exist~\cite{BGS14,FK16,LL19,FL19,Wei20c}. In particular, we present the following bi-orthogonal polynomial forms~\cite{FL19} useful for a later discussion
\begin{subequations}
	\begin{align}
		K_{00}(x,y)&=\sum_{k=0}^{m-1}p_{k}\left(x^{\theta}\right)q_{k}\left(y^{\theta}\right) ,
	\label{eq:K00op}\\
		K_{01}(x,y)&=x^{a}e^{-x}\int_{0}^{\infty}\frac{v^{a+1}e^{-v}}{x+v}K_{00}(y,v)\dd v ,
	\label{eq:K01op}\\
		K_{10}(x,y)&=y^{a+1}e^{-y}\int_{0}^{\infty}\frac{w^{a}e^{-w}}{y+w}K_{00}(w,x)\dd w ,
	\label{eq:K10op}\\
		K_{11}(x,y)&=x^{a}y^{a+1}e^{-x-y}\int_{0}^{\infty}\!\!\int_{0}^{\infty}\frac{v^{a}e^{-v}}{y+v}\frac{w^{a+1}e^{-w}}{x+w}K_{00}(v,w)\dd v\dd w-w(x,y),
	\label{eq:K11op}
	\end{align}
\end{subequations}
where the normalized bi-orthogonal polynomials
\begin{subequations}
	\begin{align}
		p_{j}\left(x\right)&=\sum_{k=0}^{j}\frac{\sqrt{2}(-1)^{k+j}\Gamma(k+j+\alpha+1)x^{k}}{\Gamma(\theta k+a+1)\Gamma(k+\alpha+1)(j-k)!k!},
	\label{eq:p}\\
		q_{j}\left(y\right)&=\sum_{k=0}^{j}\frac{\sqrt{2}(-1)^{k+j}(\theta j+a+1)\Gamma(k+j+\alpha+1)y^{k}}{\Gamma(\theta k+a+2)\Gamma(k+\alpha+1)(j-k)!k!},
	\label{eq:q}
	\end{align}
\end{subequations}
are orthogonal with respect to the $b=a+1$ specialization of the weight function \eqref{eq:wb} as given by the orthogonality condition \eqref{BOPS_norm}. Here, we are using slightly different notations of the bi-orthogonal polynomials, cf.~(\ref{hybridp}) and ~(\ref{hybridq}), which are simply related by~(\ref{eq:pp}).

Before presenting the main results on the average entropies, we provide in the following lemma a generalization of the kernel factorization property from $\theta=1$ as reported in~\cite{FK16} to an arbitrary $\theta$. This property is useful in simplifying the $k$-point densities for the higher moment calculations as demonstrated in~\cite{Wei20c,LW21} for the case $\theta=1$.
\begin{lemma}\label{l:1}
For any $\theta>0$, the correlation kernels~(\ref{eq:K00op})--(\ref{eq:K11op}) can be factorized
\begin{subequations}
	\begin{align}
			K_{00}(x,y)+K_{00}(y,x) &= u(x)u(y),
		\label{eq:ww} \\
			K_{01}(x,y)-K_{10}(y,x) &= v(x)u(y),
		\label{eq:vw} \\
			K_{11}(x,y)+K_{11}(y,x) &= -v(x)v(y),
		\label{eq:vv}
	\end{align}
\end{subequations}
as the product of the functions
\begin{subequations}
	\begin{align}
			u(x) &= \theta\sum_{k=0}^{m-1}\frac{(-1)^{k}\Gamma(k+\alpha+m+1)x^{\theta k}}{\Gamma(\theta k+a+2)\Gamma(k+\alpha+1)\Gamma(m-k)k!},
		\label{eq:wf} \\
			v(x) &= e^{-x}x^a-\theta x^{2a+1}\sum_{k=0}^{m-1}\frac{(-1)^{k}\Gamma(k+\alpha+m+1)\Gamma(-\theta k-a,x)x^{\theta k}}{(\theta k+a+1)\Gamma(k+\alpha+1)\Gamma(m-k)k!},
		\label{eq:vf}
	\end{align}
\end{subequations}
where $\Gamma(a,x)=\int_{x}^{\infty}t^{a-1}e^{-t}\dd t$ denoting the incomplete Gamma function.
\end{lemma}
\begin{proof}
The starting point of the proof is the bi-orthogonal polynomial forms of the kernels~(\ref{eq:K00op})--(\ref{eq:K11op}). To show~(\ref{eq:ww}), we first represent~(\ref{eq:K00op}) via~(\ref{eq:p}) and~(\ref{eq:q}) as
\begin{eqnarray}
	K_{00}(x,y) &=& \sum_{k=0}^{m-1}\sum_{i=0}^{m-1}\frac{\theta(-1)^{i+k}x^{\theta k}y^{\theta i}}
	{\Gamma(\theta i+a+2)\Gamma(\theta k+a+1)\Gamma(i+\alpha+1)\Gamma(k+\alpha+1)i!k!}
	\nonumber \\
	&& \times\sum_{j=i}^{m-1}(2j+\alpha+1)\frac{\Gamma(j+i+\alpha+1)\Gamma(j+k+\alpha+1)}{\Gamma(j-i+1)\Gamma(j-k+1)}
	\\
	&=& \sum_{k=0}^{m-1}\sum_{i=0}^{m-1}\frac{\theta(-1)^{i+k}x^{\theta k}y^{\theta i}}
	{\Gamma(\theta i+a+2)\Gamma(\theta k+a+1)\Gamma(i+\alpha+1)\Gamma(k+\alpha+1)i!k!}
	\nonumber \\
	&& \times\frac{\Gamma(i+\alpha+m+1)\Gamma(k+\alpha+m+1)}{(i+k+\alpha+1)\Gamma(m-i)\Gamma(m-k)},
\end{eqnarray}
where the last step is obtained by Lemma~4.1 in~\cite{BGS14}. We then have
\begin{eqnarray}
	&&K_{00}(x,y)+K_{00}(y,x) \\
	&=&\sum_{k=0}^{m-1}\sum_{i=0}^{m-1}\frac{\theta^{2}(-1)^{i+k}x^{\theta k}y^{\theta i}}
	{\Gamma(\theta i+a+2)\Gamma(\theta k+a+2)\Gamma(i+\alpha+1)\Gamma(k+\alpha+1)i!k!}
	\nonumber \\
	&& \times\frac{\Gamma(i+\alpha+m+1)\Gamma(k+\alpha+m+1)}{(i+k+\alpha+1)\Gamma(m-i)\Gamma(m-k)}\left(i+k+\frac{2(a+1)}{\theta}\right) ,
	\\
	&=& \sum_{k=0}^{m-1}\frac{\theta(-1)^{k}\Gamma(k+\alpha+m+1)x^{\theta k}}{\Gamma(\theta k+a+2)\Gamma(k+\alpha+1)\Gamma(m-k)k!}
	\nonumber \\
	&& \times\sum_{i=0}^{m-1}\frac{\theta(-1)^{i}\Gamma(i+\alpha+m+1)y^{\theta i}}{\Gamma(\theta i+a+2)\Gamma(i+\alpha+1)\Gamma(m-i)i!}=u(x)u(y),
\end{eqnarray}
which establishes~(\ref{eq:ww}). To show~(\ref{eq:vw}), we insert~(\ref{eq:K00op}) into~(\ref{eq:K01op}) and~(\ref{eq:K10op}) that gives
\begin{eqnarray*}
	&&K_{01}(x,y)-K_{10}(y,x) \\
	&=&\int_{0}^{\infty}\frac{x^{a}v^{a}e^{-x-v}}{x+v}\left(vK_{00}(y,v)-xK_{00}(v,y)\right)\dd v\\ &=&\sum_{k=0}^{m-1}\sum_{i=0}^{m-1}\frac{\theta(-1)^{i+k}\Gamma(i+\alpha+m+1)\Gamma(k+\alpha+m+1)\left(\Gamma(m-i)\Gamma(m-k)i!k!\right)^{-1}y^{\theta i}}{(i+k+\alpha+1)\Gamma(\theta i+a+2)\Gamma(\theta k+a+2)\Gamma(i+\alpha+1)\Gamma(k+\alpha+1)}\\
	&&\times\int_{0}^{\infty}\frac{x^{a}v^{a}e^{-x-v}}{x+v}v^{\theta k}\left((\theta i+a+1)v-x(\theta k+a+1)\right)\dd v\\
	&=&\sum_{i=0}^{m-1}\frac{\theta(-1)^{i}\Gamma(i+\alpha+m+1)e^{-x}x^{a}y^{\theta i}}{\Gamma(\theta i+a+1)\Gamma(i+\alpha+1)\Gamma(m-i)i!}\sum_{k=0}^{m-1}\frac{(-1)^{k}\Gamma(k+\alpha+m+1)\left(\Gamma(m-k)k!\right)^{-1}}{(i+k+\alpha+1)(\theta k+a+1)\Gamma(k+\alpha+1)}\\
	&&-\sum_{i=0}^{m-1}\frac{\theta^{2}(-1)^{i}\Gamma(i+\alpha+m+1)x^{2a+1}y^{\theta i}}{\Gamma(\theta i+a+2)\Gamma(i+\alpha+1)\Gamma(m-i)i!} \sum_{k=0}^{m-1}\frac{(-1)^{k}\Gamma(k+\alpha+m+1)\Gamma(-\theta k-a,x)x^{\theta k}}{(\theta k+a+1)\Gamma(k+\alpha+1)\Gamma(m-k)k!}\\
	&=&\left(e^{-x}x^{a}-\theta x^{2a+1}\sum_{k=0}^{m-1}\frac{(-1)^{k}\Gamma(k+\alpha+m+1)\Gamma(-\theta k-a,x)x^{\theta k}}{(\theta k+a+1)\Gamma(k+\alpha+1)\Gamma(m-k)k!}\right)\\
	&&\times\sum_{i=0}^{m-1}\frac{\theta(-1)^{i}\Gamma(i+\alpha+m+1)y^{\theta i}}{\Gamma(\theta i+a+2)\Gamma(i+\alpha+1)\Gamma(m-i)i!}=v(x)u(y),
\end{eqnarray*}
where the second to last equality is obtained by the identity
\begin{equation}\label{eq:4F3}
	\sum_{k=0}^{m-1}\frac{(-1)^{k}\Gamma(k+\alpha+m+1)}{(i+k+\alpha+1)(\theta k+a+1)\Gamma(k+\alpha+1)\Gamma(m-k)k!}=\frac{1}{\theta i+a+1}.
\end{equation}
This identity is established by the fact that the sum can be written in terms of a unit argument terminating hypergeometric function of Saalsch\"{u}tzian type~\cite{Prudnikov86} as
\begin{equation}
	{}_{4}F_{3}\left(1-m,\frac{\alpha+1}{2},\alpha+m+1,i+\alpha+1;\frac{\alpha+3}{2},\alpha+1,i+\alpha+2;1\right),
\end{equation}
which in general admits~\cite{Prudnikov86}
\begin{equation*}
	{}_{4}F_{3}(-n,b,c,d;b+1,c-l,d-m;1)=\frac{n!(b-c+1)_{l}(b-d+1)_{m}}{(b+1)_{n}(1-c)_{l}(1-d)_{m}},~~~~0\leq l+m\leq n,
\end{equation*}
with
\begin{equation}
	(a)_{n}=\frac{\Gamma(a+n)}{\Gamma(a)},
\end{equation}
denoting the Pochhammer's symbol. This completes the proof of~(\ref{eq:vw}). To show~(\ref{eq:vv}), we insert~(\ref{eq:K00op}) into~(\ref{eq:K11op}) that leads to
\begin{eqnarray*}
&&K_{11}(x,y)+K_{11}(y,x) \\
&=&x^{a}y^{a}e^{-x-y}\left(\int_{0}^{\infty}\!\!\int_{0}^{\infty}\frac{v^{a}w^{a}e^{-v-w}}{(x+w)(y+v)}\left(ywK_{00}(v,w)+xvK_{00}(w,v)\dd w\dd v\right)-1\right)\\
&=&\sum_{k=0}^{m-1}\sum_{i=0}^{m-1}\frac{\theta(-1)^{i+k}\Gamma(i+\alpha+m+1)\Gamma(k+\alpha+m+1)\left(\Gamma(m-i)\Gamma(m-k)\right)^{-1}}{(i+k+\alpha+1)\Gamma(\theta i+a+2)\Gamma(\theta k+a+1)\Gamma(i+\alpha+1)\Gamma(k+\alpha+1)i!k!}\\
&&\times x^{a}y^{a}e^{-x-y}\int_{0}^{\infty}\!\!\int_{0}^{\infty}\frac{v^{a}w^{a}e^{-v-w}\left(yv^{\theta k}w^{\theta i+1}+xw^{\theta k}v^{\theta i+1}\right)}{(x+w)(y+v)}\dd w\dd v-x^{a}y^{a}e^{-x-y}\\
&=&\sum_{k=0}^{m-1}\frac{\theta(-1)^{k}\Gamma(k+\alpha+m+1)\Gamma(-\theta k-a,x)x^{\theta k+2a+1}y^{a}e^{-y}}{\Gamma(k+\alpha+1)\Gamma(m-k)k!}\\
&&\times\sum_{i=0}^{m-1}\frac{(-1)^{i}\Gamma(i+\alpha+m+1)}{(i+k+\alpha+1)(\theta i+a+1)\Gamma(i+\alpha+1)\Gamma(m-i)i!}\\
&&+\sum_{k=0}^{m-1}\frac{\theta(-1)^{k}\Gamma(k+\alpha+m+1)\Gamma(-\theta k-a,y)y^{\theta k+2a+1}x^{a}e^{-x}}{\Gamma(k+\alpha+1)\Gamma(m-k)k!}\\
&&\times\sum_{i=0}^{m-1}\frac{(-1)^{i}\Gamma(i+\alpha+m+1)}{(i+k+\alpha+1)(\theta i+a+1)\Gamma(i+\alpha+1)\Gamma(m-i)i!}\\
&&-\theta x^{2a+1}\sum_{k=0}^{m-1}\frac{(-1)^{k}\Gamma(k+\alpha+m+1)\Gamma(-\theta k-a,x)x^{\theta k}}{(\theta k+a+1)\Gamma(k+\alpha+1)\Gamma(m-k)k!}\\
&&\times\theta y^{2a+1}\sum_{i=0}^{m-1}\frac{(-1)^{i}\Gamma(i+\alpha+m+1)\Gamma(-\theta i-a,y)y^{\theta i}}{(\theta i+a+1)\Gamma(i+\alpha+1)\Gamma(m-i)i!}-x^{a}y^{a}e^{-x-y}\\
&=&-\left(e^{-x}x^a-\theta x^{2a+1}\sum_{k=0}^{m-1}\frac{(-1)^{k}\Gamma(k+\alpha+m+1)\Gamma(-\theta k-a,x)x^{\theta k}}{(\theta k+a+1)\Gamma(k+\alpha+1)\Gamma(m-k)k!}\right)\\
&&\times\left(e^{-y}y^a-\theta y^{2a+1}\sum_{k=0}^{m-1}\frac{(-1)^{k}\Gamma(k+\alpha+m+1)\Gamma(-\theta k-a,y)y^{\theta k}}{(\theta k+a+1)\Gamma(k+\alpha+1)\Gamma(m-k)k!}\right)=-v(x)v(y),
\end{eqnarray*}
where the second to last equality is obtained by applying twice the identity~(\ref{eq:4F3}). This completes the proof of Lemma~\ref{l:1}.
\end{proof}

\subsection{Average entropies over interpolating ensemble}
With the above preparations, we now state the results on the average entropies over the interpolating ensemble. Our first result is the formula of average purity as summarized in the following proposition.
\begin{proposition}\label{p:1}
The average value of quantum purity~(\ref{eq:P}) under the quantum interpolating ensemble~(\ref{eq:QI}), for any $\theta>0$ and $a>-1$, is given by
\begin{equation}\label{eq:Pm}
\mathbb{E}_{f}\!\left[S_{\rm{P}}\right]=\sum_{k=0}^{m-1}\frac{(-1)^{k+m-1}\theta(\theta k+a+2)^2}{2d(d+1)(m-1-k)!k!}A_{\theta}(k),
\end{equation}
where
\begin{equation}\label{eq:PmA}
A_{\theta}(k)=\frac{\Gamma\left(k+\frac{2(a+2)}{\theta}\right)\Gamma\left(k+m+\frac{2(a+1)}{\theta}\right)\Gamma\left(k+1+\frac{2}{\theta}\right)}							  {\Gamma\left(k+\frac{2(a+1)}{\theta}\right)\Gamma\left(k+m+\frac{2(a+2)}{\theta}\right)\Gamma\left(k+1+\frac{2}{\theta}-m\right)}.
\end{equation}
\end{proposition}
Before proving Proposition~\ref{p:1}, two remarks on its special cases are in order.
\begin{remark}
In the special case $\theta=1$, $a=n-m-1/2$ that corresponds to the Bures-Hall ensemble~(\ref{eq:BH}), by recognizing $A_{1}(m-2)$ and $A_{1}(m-1)$ as the only non-vanishing terms in~(\ref{eq:PmA}) with
\begin{equation}
A_{1}(k)=\frac{(k+2n-2m+2)(k+2n-2m+1)\Gamma(k+3)}{(k+2n-m+2)(k+2n-m+1)\Gamma(k+3-m)},
\end{equation}
the expression~(\ref{eq:Pm}) simplifies to
\begin{align*}
	\mathbb{E}_{f}\!\left[S_{\rm{P}}\right]
	&=\sum_{k=m-2}^{m-1}\frac{2(-1)^{k+m-1}(k+n-m+3/2)^2}{m(2n-m)(2mn-m^{2}+2)(m-1-k)!k!}A_{1}(k)
\\
	&=\frac{2n(2n+m)-m^2+1}{2n\left(2mn-m^2+2\right)}.
\end{align*}
This recovers the mean purity formula of the Bures-Hall ensemble recently reported in~\cite{Sarkar19,Wei20b,LW21}.
\end{remark}
\begin{remark}
In the special case $\theta=2$, $a=n-m$ that corresponds to the Hilbert-Schmidt ensemble~(\ref{eq:HS}),
by recognizing $A_{2}(m-1)$ as the only non-vanishing term in~(\ref{eq:PmA}) with
\begin{equation}
	A_{2}(k)=\frac{(k+n-m+1)\Gamma(k+2)}{(k+n+1)\Gamma(k+2-m)},
\end{equation}
the expression~(\ref{eq:Pm}) simplifies to
\begin{align*}
	\mathbb{E}_{f}\!\left[S_{\rm{P}}\right]
	&=\sum_{k=m-1}^{m-1}\frac{(-1)^{k+m-1}(2k+n-m+2)^2}{mn(mn+1)(m-1-k)!k!}A_{2}(k)
\\
	&=\frac{m+n}{mn+1}.
\end{align*}
We recover the mean purity formula of the Hilbert-Schmidt ensemble obtained in~\cite{Lubkin78}.
\end{remark}
We now prove the Proposition~\ref{p:1}.
\begin{proof}
The essential task is to compute $\mathbb{E}_{h}\!\left[T_{\rm{P}}\right]$, which, after inserting into the moment relation~(\ref{eq:f2hP1}), will establish the Proposition~\ref{p:1}.
The required single eigenvalue density $h_{1}(x)$ of the unconstrained ensemble~(\ref{eq:uQI}) can be read off from the correlation function~(\ref{eq:rk}) that corresponds to a Pfaffian of a $2\times 2$ matrix as
\begin{equation}\label{eq:h1}
	h_{1}(x)=\frac{1}{m}\rho_{1}(x)=\frac{1}{2m}\left(K_{01}(x,x)+K_{10}(x,x)\right).
\end{equation}
The computation now boils down to computing two integrals
\begin{align}
	\mathbb{E}_{h}\!\left[T_{\rm{P}}\right]
	&= m\int_{0}^{\infty}x^{2}h_{1}(x)\dd x
\nonumber \\
	&=\frac{1}{2}\int_{0}^{\infty}x^{2}K_{01}(x,x)\dd x+\frac{1}{2}\int_{0}^{\infty}x^{2}K_{10}(x,x)\dd x.
\label{eq:2IP}
\end{align}
The starting point to calculate the above integrals is the fact that the contour form~(\ref{eq:FH}) of the Fox H-function~(\ref{eq:H1}) admits a finite number of single poles, which by residue calculation gives a finite sum
\begin{equation}\label{eq:FHs}
	H_{q}(x)=\sum_{k=0}^{m-1}\frac{(-1)^{k}\Gamma(k+\alpha+m+1)\left(tx^{\theta}\right)^k}{\Gamma(k+\alpha+1)\Gamma(\theta k+q+1)(m-1-k)!k!}.
\end{equation}
Therefore, we have
\begin{eqnarray}
	&&\int_{0}^{\infty}x^{2}K_{01}(x,x)\dd x \\
	&=&\int_{0}^{\infty}\theta x^{2a+3}\int_{0}^{1}t^{\alpha}H_{a}(y)G_{a+1}(x)\dd t\dd x\\
	&=&\sum_{k=0}^{m-1}\frac{\theta(-1)^{k}\Gamma(k+\alpha+m+1)}{\Gamma(k+\alpha+1)\Gamma(\theta k+a+1)(m-1-k)!k!}\int_{0}^{\infty}\!\!x^{\theta k+2a+3}\int_{0}^{1}\!t^{\alpha+k}G_{a+1}(x)\dd t\dd x
\label{eq:dtdx} \\
	&=&\sum_{k=0}^{m-1}\frac{(-1)^{k+m-1}\theta(\theta k+a+1)(\theta k+a+2)}{2(m-1-k)!k!}A_{\theta}(k)
\label{eq:2IP1}
\end{eqnarray}
where the integrals over $t$ and $x$ in~(\ref{eq:dtdx}) are evaluated respectively by the identity~\cite{Mathai10}
\begin{eqnarray}\label{eq:tI}
&&\int_{0}^{1}x^{\rho-1}H_{p,q}^{m,n}\left(\begin{array}{c}(a_{1},A_{1}),\ldots,(a_{n},A_{n});(a_{n+1},A_{n+1}),\ldots,(a_{p},A_{p})\\ (b_{1},B_{1}),\ldots,(b_{m},B_{m});(b_{m+1},B_{m+1}),\ldots,(b_{q},B_{q})\end{array}\Big|\eta x\Big.\right)\dd x\\	&=&H_{p+1,q+1}^{m,n+1}\left(\begin{array}{c}(1-\rho,1),(a_{1},A_{1}),\ldots,(a_{n},A_{n});(a_{n+1},A_{n+1}),\ldots,(a_{p},A_{p})\\(b_{1},B_{1}),\ldots, (b_{m},B_{m});(b_{m+1},B_{m+1}),\ldots,(b_{q},B_{q}),(-\rho,1)\end{array}\Big|\eta\Big.\right) ,
\end{eqnarray}
and the Mellin transform of the Fox H-function~\cite{Mathai10}, cf.~(\ref{eq:FH}),
\begin{eqnarray}\label{eq:xI}
&&\int_{0}^{\infty}x^{s-1}H_{p,q}^{m,n}\left(\begin{array}{c}(a_{1},A_{1}),\ldots,(a_{n},A_{n});(a_{n+1},A_{n+1}),\ldots,(a_{p},A_{p})\\ (b_{1},B_{1}),\ldots,(b_{m},B_{m});(b_{m+1},B_{m+1}),\ldots,(b_{q},B_{q})\end{array}\Big|\eta x\Big.\right)\dd x\\
&=&{\frac{\eta^{-s}\prod_{j=1}^m\Gamma\left(b_{j}+B_{j}s\right)\prod_{j=1}^n\Gamma\left(1-a_{j}-A_{j}s\right)}{\prod_{j=n+1}^p \Gamma\left(a_{j}+A_{j}s\right)\prod_{j=m+1}^q\Gamma\left(1-b_{j}-B_{j}s\right)}}.
\end{eqnarray}
In the same manner, the second integral in~(\ref{eq:2IP}) is evaluated to
\begin{equation}\label{eq:2IP2}
	\int_{0}^{\infty}x^{2}K_{10}(x,x)\dd x=\sum_{k=0}^{m-1}\frac{(-1)^{k+m-1}\theta(\theta k+a+2)(\theta k+a+3)}{2(m-1-k)!k!}A_{\theta}(k).
\end{equation}
Putting together~(\ref{eq:f2hP1}),~(\ref{eq:2IP}),~(\ref{eq:2IP1}), and~(\ref{eq:2IP2}), we complete the proof of Proposition~\ref{p:1}. \hspace*{\fill}
\end{proof}
The next result on the average von Neumann entropy is presented in the following proposition.
\begin{proposition}\label{p:2}
The average value of von Neumann entropy~(\ref{eq:vN}) under the quantum interpolating ensemble~(\ref{eq:QI}), for any $\theta>0$ and $a>-1$, is given by
\begin{equation}\label{eq:vNm}
	\mathbb{E}_{f}\!\left[S_{\rm{vN}}\right]=\psi_{0}(d+1)-\sum_{k=0}^{m-1}\frac{(-1)^{k+m-1}\left(\theta k+a+3/2\right)}{d(m-1-k)!k!}B_{\theta}(k),
\end{equation}
where
\begin{eqnarray}
B_{\theta}(k)&=&\frac{\Gamma\left(k+\frac{2a+3}{\theta}\right)\Gamma\left(k+m+\frac{2a+2}{\theta}\right)\Gamma\left(k+1+\frac{1}{\theta}\right)}{\Gamma\left(k+\frac{2a+2}{\theta}\right)\Gamma\left(k+m+\frac{2a+3}{\theta}\right)\Gamma\left(k+1+\frac{1}{\theta}-m\right)}\Bigg(\psi_{0}\left(k+\frac{2a+3}{\theta}\right)\nonumber\\
&&+\psi_{0}\left(k+1+\frac{1}{\theta}\right)-\psi_{0}\left(k+m+\frac{2a+3}{\theta}\right)-\psi_{0}\left(k+1+\frac{1}{\theta}-m\right)\nonumber\\
&&+\theta\left(\psi_{0}(\theta k+a+2)-\frac{\theta k+a+1}{\theta k+a+3/2}\right)\!\Bigg).
\end{eqnarray}
\end{proposition}
	
\begin{proof}
The main task is to compute the average $\mathbb{E}_{h}\!\left[T_{\rm{vN}}\right]$, which, after inserting into the moment relation~(\ref{eq:f2hvN1}), establishes Proposition~\ref{p:2}.
By employing the single eigenvalue density~(\ref{eq:h1}), this task boils down to computing two integrals
\begin{equation}\label{eq:2IvN}
	\mathbb{E}_{h}\!\left[T_{\rm{vN}}\right]=\frac{1}{2}\int_{0}^{\infty}x\ln xK_{01}(x,x)\dd x+\frac{1}{2}\int_{0}^{\infty}x\ln xK_{10}(x,x)\dd x.
\end{equation}
We first compute the integral
\begin{equation}
	\int_{0}^{\infty}x^{\beta}K_{01}(x,x)\dd x,\quad \beta>0,
\end{equation}
by using the results~(\ref{eq:FHs}),~(\ref{eq:tI}), and~(\ref{eq:xI}) as
\begin{align}
	\int_{0}^{\infty}x^{\beta}K_{01}(x,x)\dd x
	=&\sum_{k=0}^{m-1}\frac{(-1)^{k+m}\Gamma\left(k+\alpha+m+1\right)}{2\theta(m-1-k)!k!\Gamma\left(k+\alpha+1\right)\Gamma(\theta k+a+1)}
	\nonumber\\
	&\times\frac{\Gamma(s)\Gamma(s-\alpha)\Gamma(k+\alpha-s+1)\Gamma(\theta s-a-1)}{\Gamma(s+m)\Gamma(s-\alpha-m)\Gamma(k+\alpha-s+2)},\label{eq:xb01}
\end{align}
where $\alpha$ is given by~(\ref{eq:alpha}) and we denote
\begin{equation}
	s=\beta-1+k+\frac{2a+3}{\theta}.
\end{equation}
Similarly, we also obtain
\begin{align}
	\int_{0}^{\infty}x^{\beta}K_{10}(x,x)\dd x
	=&\sum_{k=0}^{m-1}\frac{(-1)^{k+m}\Gamma\left(k+\alpha+m+1\right)}{2\theta(m-1-k)!k!\Gamma\left(k+\alpha+1\right)\Gamma(\theta k+a+2)}
	\nonumber\\
	 &\times\frac{\Gamma(s)\Gamma(s-\alpha)\Gamma(k+\alpha-s+1)\Gamma(\theta s-a)}{\Gamma(s+m)\Gamma(s-\alpha-m)\Gamma(k+\alpha-s+2)}. \label{eq:xb10}
\end{align}
Taking the derivative of~(\ref{eq:xb01}) and~(\ref{eq:xb10}) with respect to $\beta$ before setting $\beta\to1$ leads to the desired expression for~(\ref{eq:2IvN}), which upon inserting into the moment relation~(\ref{eq:f2hvN1}) completes the proof of Proposition~\ref{p:2}.
\hspace*{\fill}
\end{proof}

\begin{figure}[t!]
	\centering
	\includegraphics[width=0.64\linewidth]{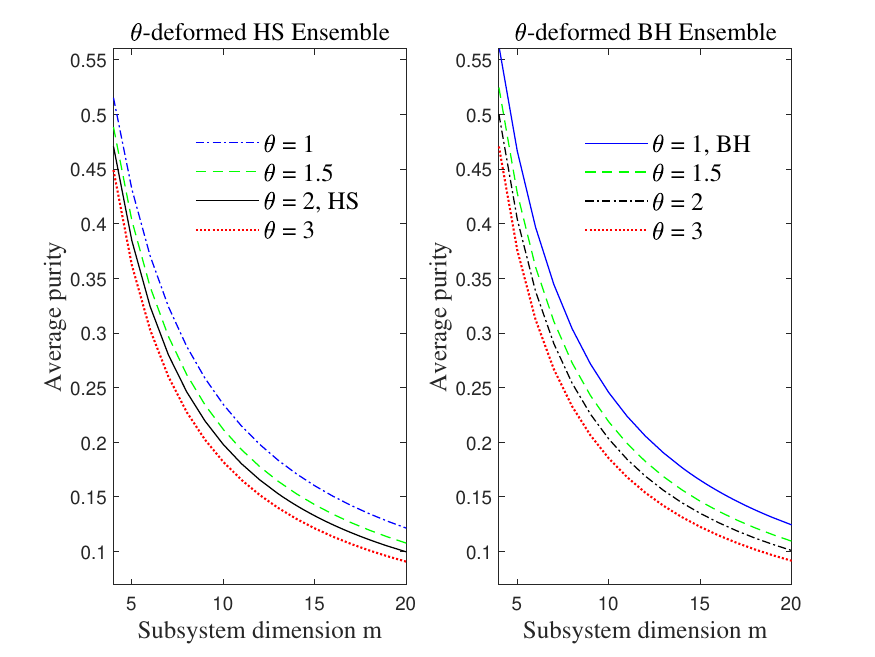}
	\caption{Average purity~(\ref{eq:Pm}) as a function of subsystem dimensions: the impact of $\theta$-deformation. The two solid curves represent the cases of the standard ensembles~(\ref{cases}) with no deformations and the other curves represent the cases of the deformed ensemble~(\ref{eq:QI}). In all cases, we consider equal subsystem dimensions $m=n$.}
	\label{fig:Fig1}
\end{figure}

\begin{remark}
In the special case $\theta=1$, $a=n-m-1/2$ that corresponds to the Bures-Hall ensemble~(\ref{eq:BH}), the result~(\ref{eq:vNm}) in Proposition~\ref{p:2} reduces to
\begin{align*}
	\mathbb{E}_{f}\!\left[S_{\rm{vN}}\right]
	=&\psi_{0}\left(mn-\frac{m^2}{2}+1\right)-\psi_{0}\left(n+\frac{1}{2}\right)+\psi_{0}(2n+1)-\psi_{0}(2n-m+1)+\psi_{0}(1)
\\
	 &-\psi_{0}(m+1)+\frac{2n-1}{2n}+\sum_{k=0}^{m-2}\frac{2(k+1)(k+n-m+1)(k+2n-2m+1)}{m(2n-m)(m-1-k)(k+2n-m+1)}
\\
	=&\psi_{0}\left(mn-\frac{m^2}{2}+1\right)-\psi_{0}\left(n+\frac{1}{2}\right).
\end{align*}
This recovers the mean formula of von Neumann entropy under the Bures-Hall ensemble recently studied in~\cite{Sarkar19,Wei20b}.
\end{remark}
\begin{remark}
In the special case $\theta=2$, $a=n-m$ that corresponds to the Hilbert-Schmidt ensemble~(\ref{eq:HS}), the mean formula of von Neumann entropy is well-known
\begin{equation}
	\mathbb{E}_{f}\!\left[S_{\rm{vN}}\right]=\psi_{0}(mn+1)-\psi_{0}(n)-\frac{m+1}{2n},
\end{equation}
which was conjectured by Page~\cite{Page93} and later proved in~\cite{Foong94,Ruiz95}. By equating Proposition~\ref{p:2} in this special case to the above result of Page, one arrives at the following non-trivial summation identity
\begin{eqnarray} &&\frac{1}{mn}\sum_{k=0}^{m-1}\frac{(-1)^{k+m-1}\left(2k+n-m+\frac{3}{2}\right)\Gamma\left(k+\frac{3}{2}\right)\Gamma(k+n+1)\Gamma\left(k+n-m+\frac{3}{2}\right)}
{(m-1-k)!k!\Gamma\left(k-m+\frac{3}{2}\right)\Gamma\left(k+n+\frac{3}{2}\right)\Gamma(k+n-m+1)}\nonumber\\
&&\times\Bigg(\psi_{0}\left(k+\frac{3}{2}\right)+\psi_{0}\left(k+n-m+\frac{3}{2}\right)-\psi_{0}\left(k+n+\frac{3}{2}\right)-\psi_{0}\left(k-m+\frac{3}{2}\right)\nonumber\\
&&+2\psi_{0}(2k+n-m+2)-\frac{2(2k+n-m+1)}{2k+n-m+3/2}\Bigg)=\psi_{0}(n)+\frac{m+1}{2n},
\end{eqnarray}
a direct proof of which, however, seems difficult.
\end{remark}

\begin{figure}[t!]
	\centering
	\includegraphics[width=0.64\linewidth]{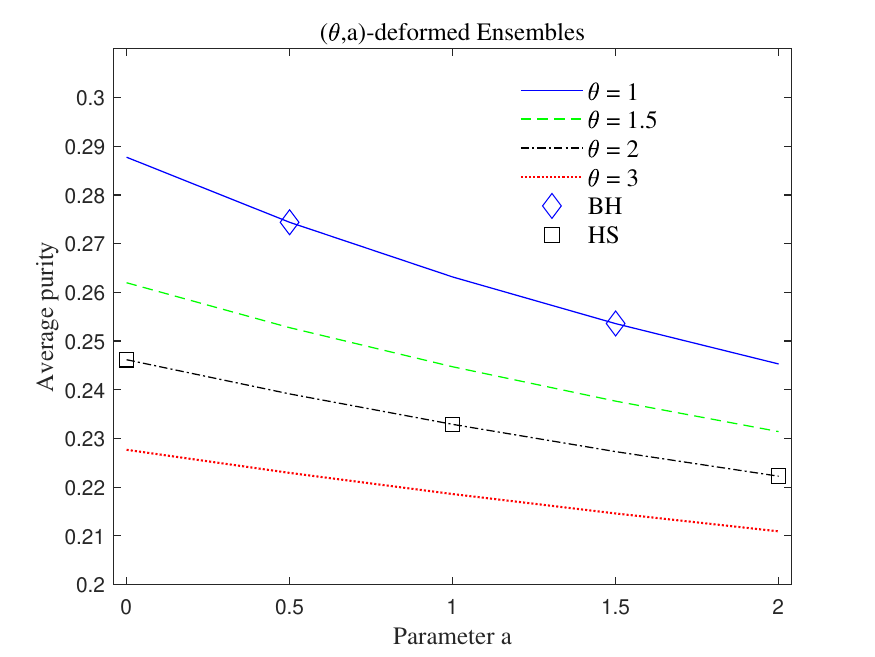}
	\caption{Average purity~(\ref{eq:Pm}) as a function of the parameters $a$ and $\theta$. The data points marked by diamond and square shapes represent the special cases of the undeformed ensembles~(\ref{cases}). In all cases, the subsystem dimension is $m=8$.}
	\label{fig:Fig2}
\end{figure}

\begin{figure}[t!]
	\centering
	\includegraphics[width=0.64\linewidth]{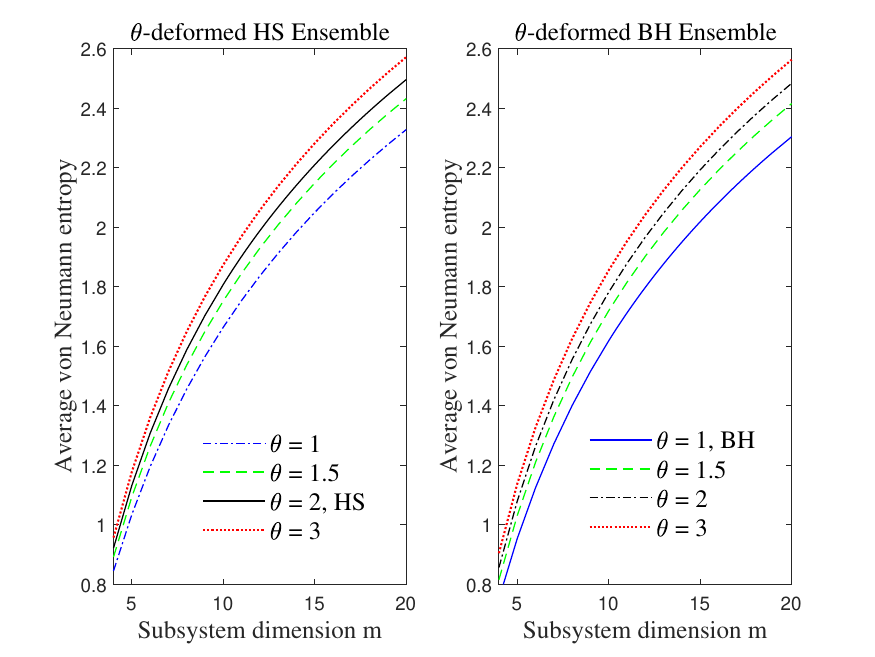}
	\caption{Average von Neumann entropy~(\ref{eq:vNm}) as a function of subsystem dimensions: the impact of $\theta$-deformation. The two solid curves represent the cases of the standard ensembles~(\ref{cases}) with no deformations and the other curves represent the cases of the deformed ensemble~(\ref{eq:QI}). In all cases, we consider equal subsystem dimensions $m=n$.}
	\label{fig:Fig3}
\end{figure}

\begin{figure}[t!]
	\centering
	\includegraphics[width=0.64\linewidth]{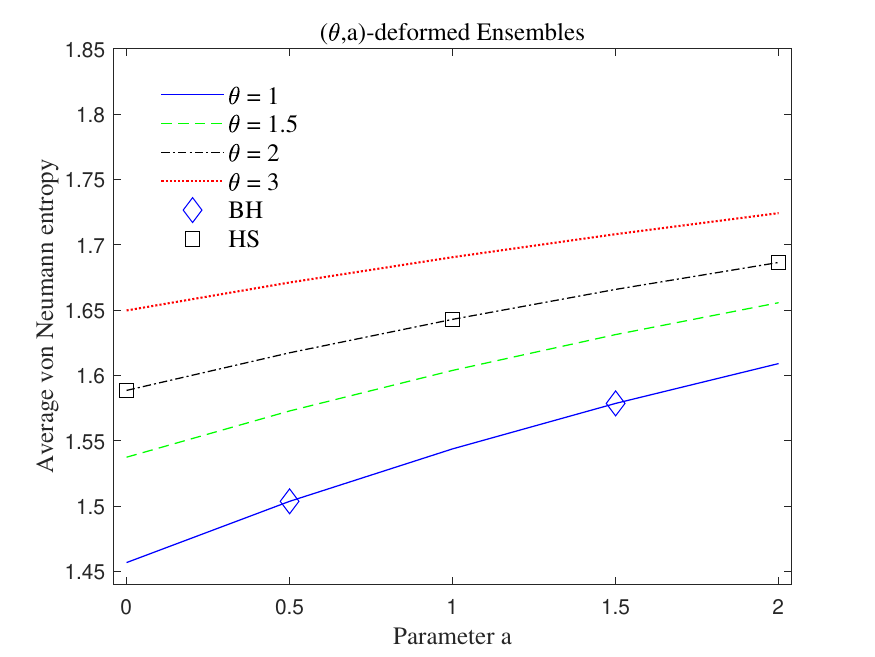}
	\caption{Average von Neumann entropy~(\ref{eq:vNm}) as a function of the parameters $a$ and $\theta$. The data points marked by diamond and square shapes represent the special cases of the undeformed ensembles~(\ref{cases}). In all cases, the subsystem dimension is $m=8$.}
	\label{fig:Fig4}
\end{figure}

\subsection{Numerical results}
We now perform some numerical studies of the average entanglement entropies over the interpolating ensemble. We first focus on the result of average purity in Proposition~\ref{p:1}. In Figure~\ref{fig:Fig1}, we plot the numerical values of average purity~(\ref{eq:Pm}) as a function of the subsystem dimensions. For different values of $\theta$, we consider both the $\theta$-deformed Hilbert-Schmidt ensemble assuming $a=0$ and and the $\theta$-deformed Bures-Hall ensemble assuming $a=-1/2$ as shown in the left subfigure and the right subfigure, respectively. Note that the choices of $a$ imply equal subsystem dimensions $m=n$ in both cases. The solid curves in Figure~\ref{fig:Fig1} describe the behavior of the standard ensembles~(\ref{cases}) with no deformations, whereas the other curves represent the corresponding $\theta$-deformed ones. It is observed that as the deformation parameter $\theta$ increases, the values of average purity decrease monotonically resulting in estimations of entanglement towards more entangled states. The observation suggests that the proposed interpolating ensemble~(\ref{eq:QI}) is indeed able to continuously interpolate among the possible values of purity by varying the $\theta$ parameter. It is also observed in Figure~\ref{fig:Fig1} that for a given $\theta$ the average purity under the Bures-Hall ensemble tends to an estimate of more separable state (i.e., a larger purity value) than that of the Hilbert-Schmidt ensemble. On the other hand, the differences are diminishing as the dimension increases. This behavior has also been recently observed in~\cite{Sarkar19}.

Since the parameter $a$ of the proposed ensemble~(\ref{eq:QI}) can be also considered as a deformation variable, we wish to understand its impact on the quantum purity. In Figure~\ref{fig:Fig2}, we plot the average purity~(\ref{eq:Pm}) as a function of the parameter $a$ for different values of $\theta$, where the dimension of subsystem is assumed to be $m=8$. The data points marked by diamond shape for $\theta=1$ and square shape for $\theta=2$ corresponds to the special case of Bures-Hall ensemble and Hilbert-Schmidt ensemble, respectively. It is observed in Figure~\ref{fig:Fig2} that as $a$ increases, the values of average purity decrease monotonically indicating more entangled states. In particular, for the cases $\theta=1$ and $\theta=2$, the average purity is seen to interpolate continuously among the permissible values~(\ref{cases}) of the parameter $a$.

We now turn to the numerical study of the von Neumann entropy in Proposition~\ref{p:2}. In Figure~\ref{fig:Fig3}, we plot the average von Neumann entropy~(\ref{eq:vNm}) as a function of the subsystem dimensions for different values of $\theta$. We consider both the $\theta$-deformed Hilbert-Schmidt and Bures-Hall ensembles with the same values of $a$ as in Figure~\ref{fig:Fig1}. It is seen that as the deformation parameter $\theta$ increases, the average von Neumann entropy increases monotonically, which also results in estimations of entanglement towards more entangled states as in Figure~\ref{fig:Fig1}. In particular, the proposed interpolating ensemble~(\ref{eq:QI}) continuously interpolates among the possible values of the von Neumann entropy. Similar to Figure~\ref{fig:Fig1}, we also observe in Figure~\ref{fig:Fig3} that the average von Neumann entropy under the Bures-Hall ensemble tends to an estimate of more separable state (i.e., a smaller value of von Neumann entropy) than that of the Hilbert-Schmidt ensemble. The differences, however, diminish as the dimension increases, which is in line with the recent observation~\cite{Sarkar19}. To understand the impact of parameter $a$, we plot in Figure~\ref{fig:Fig4} the average von Neumann entropy~(\ref{eq:vNm}) as a function of the parameter $a$ for different values of $\theta$. The dimension of subsystem is also assumed to be $m=8$. Similarly as observed in Figure~\ref{fig:Fig2}, the values of average von Neumann entropy increase monotonically indicating more entangled states as $a$ increases. Finally, we point out that various other numerical simulations have been performed, where the same relative behavior as discussed in above four figures persists.
	
\section{Conclusions}
In this work, we proposed and studied a generalized ensemble that interpolates between the two major measures of density matrices - the Hilbert-Schmidt ensemble and the Bures-Hall ensemble. In particular, we derived recurrence relations of the underlying bi-orthogonal polynomials of the ensemble useful in computing different statistical quantities. As an application, we computed the average entanglement entropies over the interpolating ensemble generalizing various known results in the literature. Numerical simulations show that the proposed ensemble provides additional power in estimating the degree of entanglement by varying the deformation parameters. Future work includes further study of the statistical information of the ensemble such as higher order moments of entropies, fidelity, and volumes as well as further study of the associated bi-orthogonal system.

\section*{Acknowledgments}
We wish to thank Shi-Hao Li for correspondence. The work of Lu Wei is supported in part by the U.S. National Science Foundation ($\#$2150486).

\end{document}